\documentclass[11pt]{article}

\usepackage{cite}
 \usepackage{subfigure}
\usepackage{multirow}
\usepackage{helvet}
\usepackage{amsmath}
\usepackage{amssymb}
\usepackage{setspace}
\usepackage{graphicx}
\usepackage{empheq}
\usepackage{soul}
\usepackage{tabularx}
\usepackage{array}
\usepackage{float}
\usepackage{braket}
\usepackage[utf8]{inputenc}
\usepackage{url}
\usepackage{hyperref}
\usepackage{slashed}
\usepackage[font=footnotesize,labelfont=bf]{caption}
\usepackage{color}
\usepackage{cleveref}

\def\be{\begin{equation}}
\def\ee{\end{equation}}

\newcommand{\anlo}{aN${}^3$LO }

\newcommand{\as}{\alpha_{S}(M_Z^2)}
\usepackage[a4paper,top=3cm,bottom=3cm,left=2.5cm,right=2.5cm,bindingoffset=0mm]{geometry}

\begin{document}
\hspace*{\fill} DESY-24-044

\begin{center}

{\Large \bf A first determination of the strong coupling $\alpha_S$ at  \\ \vspace{2mm} approximate ${\rm N}^3$LO order in a global PDF fit}

\vspace*{1cm}
T. Cridge$^a$, L. A. Harland-Lang$^{b}$, 
and R.S. Thorne$^b$\\                                               
\vspace*{0.5cm}                                                    
   
$^a$  Deutsches Elektronen-Synchrotron DESY, Notkestr. 85, 22607 Hamburg, Germany   \\  
$^b$ Department of Physics and Astronomy, University College London, London, WC1E 6BT, UK     

\begin{abstract}
\noindent We present the first determination of the value of the strong coupling via a simultaneous global fit of the proton parton distribution functions (PDFs) at approximate ${\rm N}^3$LO (a${\rm N}^3$LO) order in QCD. This makes use of the MSHT global PDF fitting framework, and in particular the recent theoretical advances that allow a PDF fit to now be performed at this order. The value of the strong coupling is found to be $\as({\rm aN^3LO}) = 0.1170 \pm 0.0016$. This is in excellent agreement with the NNLO value of $ \as({\rm NNLO}) = 0.1171 \pm 0.0014$, indicating that good perturbative convergence has been found. The resulting uncertainties, calculated using the MSHT dynamic tolerance procedure, are somewhat larger, but more accurate, at a${\rm N}^3$LO, due to the missing higher order theoretical uncertainties that are included at this order, but not at NNLO. We in addition present a detailed breakdown of the individual dataset sensitivity to the value of the strong coupling, with special focus on the impact of fitting dijet rather than inclusive jet data. This choice is found to have a non--negligible impact, but with overall good consistency found, especially at a${\rm N}^3$LO.

\end{abstract}

\end{center}

\section{Introduction}

There has been a huge amount of progress in the calculation of higher-order corrections to processes in QCD in recent years. A very large number of final states are now known exactly at NNLO, and combined with the required splitting functions and transition matrix elements for flavour thresholds this has allowed the evolution from somewhat approximate NNLO determinations of parton distribution functions (PDFs) to NNLO determinations with only a very small number of rather minor approximations~\cite{Bailey:2020ooq,Hou_2021,NNPDF:2021njg, ATLAS:2021vod,ABMP16}. The state of N$^3$LO calculations in 
perturbative QCD is now similar to that of NNLO calculations about 20 years ago, 
with light flavour structure function coefficient functions known exactly~\cite{Vermaseren:2005qc}, 
Drell Yan cross sections known to a large extent (though beginning the transition to becoming usable in PDF fits) and a great deal known about splitting 
functions~\cite{Moch:2017uml,Moch:2021qrk,Falcioni:2023luc,Falcioni:2023vqq,Moch:2023tdj,Falcioni:2023tzp,Gehrmann:2023cqm,Fadin:1975cb,Kuraev:1976ge,Lipatov:1976zz,Kuraev:1977fs,Fadin:1998py,Jaroszewicz:1982gr,Ciafaloni:1998gs,Catani:1994sq,Davies:2022ofz} particularly nonsinglet, even if some uncertainty remains. Similarly, there has been much work on heavy flavour transition matrix elements~\cite{Kawamura:2012cr,Bierenbaum:2009mv,Ablinger:2014vwa,Ablinger:2014nga,Blumlein:2021enk,Ablinger:2014uka,Ablinger:2014tla,ablinger:agq,Ablinger:2022wbb,Ablinger:2023ahe,Ablinger:2024xtt}. On this basis we recently made use of information so far available to determine for the first time a set of approximate N$^3$LO PDFs \cite{McGowan:2022nag} (this has  recently also been done in \cite{NNPDF:2024nan}), and as a by product, by investigating the possible uncertainty in the N$^3$LO extraction also obtained the dominant theoretical uncertainty on the PDFs.

Of course, as well as the input PDFs and the cross sections for final states, a 
global PDF fit also relies on the strong coupling constant, $\alpha_S$. The evolution of the coupling at N$^3$LO has been known for many years \cite{vanRitbergen:1997va}, so is taken account of fully in \cite{McGowan:2022nag}, and indeed a brief presentation of the best fit value and the variation with $\alpha_S(M_Z^2)$ was given in Section 8.9. However, we did not make a full study of the uncertainty within our fitting framework. In this article we update this study, with some very minor modifications to the central analysis (as detailed in \cite{Cridge:2023ozx,Cridge:2023ryv}), but more 
importantly, with a full study of the uncertainty. In many respects this is similar to the result found at NNLO \cite{Cridge:2021qfd}, but with a slightly enlarged, and improved uncertainty due to the inclusion of theoretical uncertainties. Moreover, we very recently considered the impact of using dijet data rather than inclusive jet data in the MSHT PDF fit \cite{Cridge:2023ozx}. Since jet data provides a significant constraint on $\alpha_S$ we consider how it affects the results of the determination of $\alpha_S(M_Z^2)$ within the global fit at both NNLO and aN$^3$LO. We find consistency between results with inclusive jet or dijet data at both NNLO and aN$^3$LO, but with improved agreement at aN$^3$LO, where the results are almost identical. 

Hence, we present our full analysis using the current default of inclusive jet data in Section 2, highlighting also the most constraining data sets at aN$^3$LO. In Section 3 we discuss the sensitivity of our results to choices made in our study, first briefly discussing the sensitivity related to the uncertainty in the splitting functions, and then comparing in detail the results using inclusive jets or dijet data. In Section 4 we show how the PDFs change with the value of $\alpha_S(M_Z^2)$ and also compare the PDF and correlated $\alpha_S$ uncertainty on  some benchmark cross sections, showing a small increase in uncertainty at aN$^3$LO due to the correct inclusion of a theoretical uncertainty.  Finally, in Section~\ref{sec:summary} we present  the conclusions and outlook for this study.

\section{Strong Coupling Dependence at NNLO and \anlo}

The baseline dataset and theoretical ingredients for our study are as presented in~\cite{Cridge:2023ozx}. That is, we now include the ATLAS 8 TeV jet data~\cite{ATLAS:2017kux}, while the treatment of the CMS inclusive jet and DY data is updated, again as described in~\cite{Cridge:2023ozx}. We in particular do not update our results to account for the most recent theoretical calculations of the splitting functions and transition matrix elements at N${}^3$LO~\cite{Falcioni:2023luc,Falcioni:2023vqq,Falcioni:2023tzp,Moch:2023tdj,Ablinger:2022wbb,Gehrmann:2023cqm,Ablinger:2023ahe,Ablinger:2024xtt}. This is in part to maintain consistency with the original  analysis of~\cite{McGowan:2022nag}, but also as a full update to account for these has been postponed until all major relevant ingredients are available. The calculation of the $Hg$ transition matrix element in~\cite{Ablinger:2024xtt} has now appeared, in the final stages of the preparation of this manuscript, and a full update will be performed in an upcoming publication. We have nonetheless checked that running with the updated splitting functions of \cite{Falcioni:2023luc,Falcioni:2023vqq,Falcioni:2023tzp,Moch:2023tdj}, which we would expect to contain the most significant dependence of the theoretical ingredients on $\as$ (as explained later in Section~\ref{subsec:sensitivity_splits}), we obtain a best fit value of $\as$ at \anlo consistent with our results in Section~\ref{subsec:bestfit} and well within the $\as$ bounds given in Section~\ref{subsec:bounds}. In fact the value obtained is very close to the minimum we obtain with our versions of the approximations to the splitting functions at N$^{3}$LO, providing further support for the reliability of the original procedure.

\subsection{Best fit $\alpha_S(M_Z^2)$ value} \label{subsec:bestfit}

The default PDFs provided in the MSHT20nnlo \cite{Bailey:2020ooq} and MSHT20an3lo \cite{McGowan:2022nag} sets are given at fixed $\as$=$0.118$, equal to the Particle Data Group value \cite{Huston:2023ofk}. However, the PDFs themselves are sensitive to the value of the strong coupling through the coefficient functions and PDF evolution. As a result we may also allow the value of the strong coupling to be free in the  fit and hence extract a best fit value of $\as$ with corresponding uncertainties. 

\begin{figure}
\begin{center}
\includegraphics[scale=0.21]{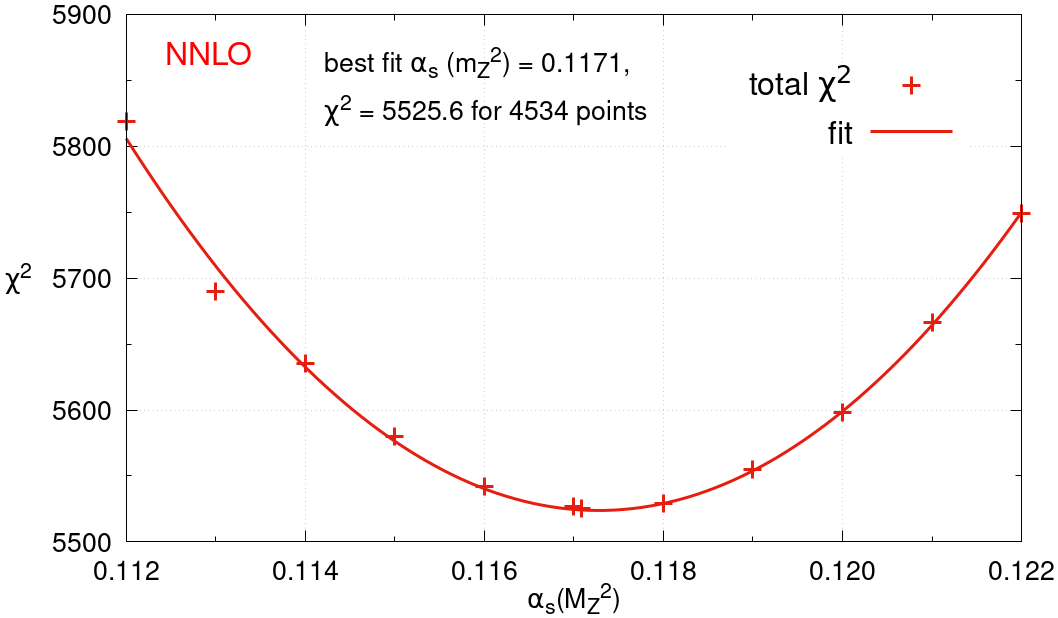}
\includegraphics[scale=0.21]{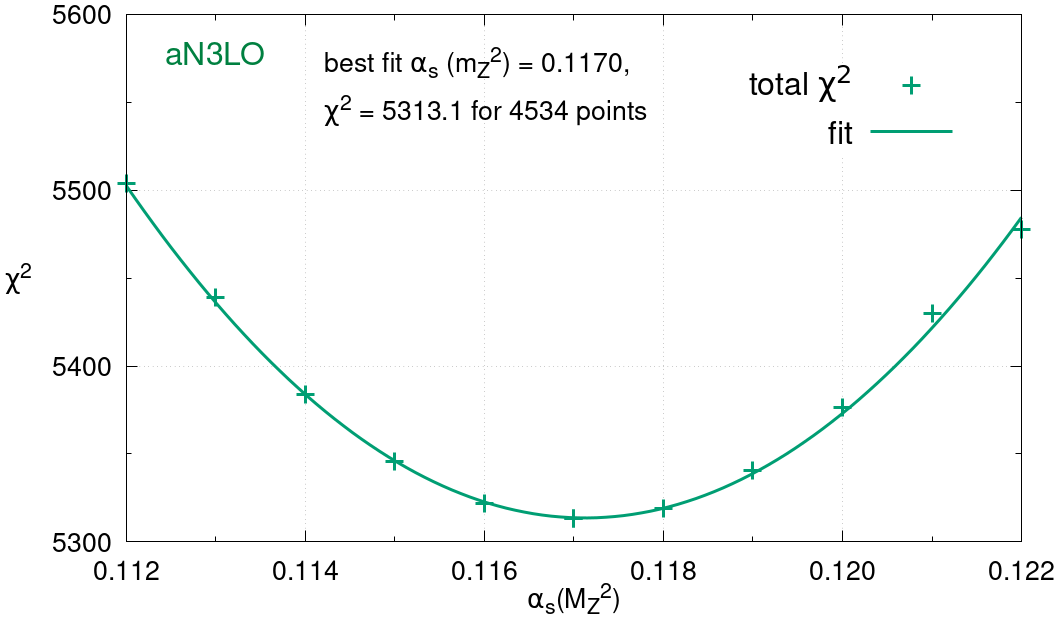}
\caption{\sf The $\chi^2$ profile for the NNLO (left) and \anlo (right) PDF fits as $\alpha_S(M_Z^2)$ is scanned from 0.112 to 0.122. The best fit $\alpha_S(M_Z^2)$ value, best fit $\chi^2$ and number of datapoints are given on the figures.}
\label{fig:chi2_profiles_total_NNLOaN3LO}
\end{center}
\end{figure}

Doing this at NNLO and \anlo we obtain respectively 0.1171 and 0.1170. The NNLO value is consistent with our previous best fit of 0.1174$\pm$0.0013 \cite{Cridge:2021qfd,dEnterria:2022hzv}. Considering that the associated NLO best fit $\as$ obtained in this previous study was 0.1203$\pm$0.0015, we observe improved perturbative convergence between the NNLO and \anlo determinations. In addition to determining the best fit, one may also scan $\as$, refitting the PDFs at each step to obtain the $\chi^2$ profile of the global PDF fit with $\as$. The corresponding profiles at NNLO (left) and \anlo (right) are shown in Fig.~\ref{fig:chi2_profiles_total_NNLOaN3LO}. The points represent the fit qualities in $\chi^2$ of the individual $\as$ fits whilst the line represents a quadratic fit, demonstrating the expected quadratic behaviour of the profiles about their global minima and indeed across the whole large range of $\as$ considered. 

The $\chi^2$ values of the global minima are given on the figures at NNLO and aN$^{3}$LO. In addition we provide in Table~\ref{tab:alphas_deltachisqs} the changes in the global PDF fit $\chi^2$ as $\as$ is scanned across a wide range of values. As well as the result for the  baseline fit we also show the corresponding values for a fit including dijet data instead, as described further in Section~\ref{sec:dijets}.  This information may be of use for analyses which wish to extract $\as$ from individual measurements, while still bearing in mind the variation of the PDF global fit quality with $\as$. Nonetheless, the correlations between the new measurement and the data in the PDF fit would still neglected in this case and the issues raised by \cite{Forte:2020pyp} for individual $\as$ extractions still apply. For these reasons, the determination of $\as$ in a global fit simultaneously with the PDFs (i.e. refitting the PDFs with $\as$) remains necessary and of significant interest.

The $\chi^2$ values at the best fit $\as$ demonstrate the preference in the fit for aN$^{3}$LO, with the best fit at \anlo approximately 212 points better in $\chi^2$ for the same number of datapoints. This comes at the expense of only 20 additional theory ``nuisance'' parameters included in the \anlo fit with respect to the NNLO, parameterising the missing pieces of the N$^{3}$LO information and hence the missing higher order uncertainties (MHOUs) in the fit. This improvement is similar to, though somewhat larger than, that observed at the default value of $\as$=$0.118$ with which the global PDFs are provided. 
This difference can be explained by the fact that the comparison is being made at slightly different values of $\as$, corresponding to the best fits at NNLO and aN$^{3}$LO, as well as the addition of the ATLAS 8~TeV inclusive jet data, and the other theoretical updates described in~\cite{Cridge:2023ozx}. We observe that the Deep Inelastic Scattering (DIS) datasets, for which the N$^{3}$LO theory is almost complete, i.e. the light quark coefficient functions are known exactly \cite{Vermaseren:2005qc}, change by $\Delta\chi^2=-97$ in going from NNLO to aN$^{3}$LO, whilst Drell-Yan changes by $\Delta \chi^2=-25$ (with a negative value corresponding to an improvement in $\chi^2$). The inclusive jets, vector boson plus jets (including $Z$ $p_T$ data), top production, and Semi-inclusive DIS datasets all also improve, changing by $\Delta\chi^2=-12,-76,-3, -5$ respectively. These are similar to the observations in \cite{McGowan:2022nag}, with the DIS and vector boson plus jets data providing the most significant improvements in going from NNLO to aN$^{3}$LO. Nonetheless, we now observe somewhat greater improvements in the Drell-Yan, SIDIS and inclusive jets data, with the latter a direct impact of the additional inclusion of the ATLAS 8~TeV inclusive jets data, as reported in \cite{Cridge:2023ozx}.

\begin{center}
\begin{table}[htbp!]
\fontsize{9.5}{12}\selectfont 
 \renewcommand\arraystretch{0.99} 
\centering \hspace{-0.1cm} \begin{tabular}{|>{\centering\arraybackslash}m{1.75cm}|>{\centering\arraybackslash}m{2.75cm}|>{\centering\arraybackslash}m{2.75cm}|>{\centering\arraybackslash}m{2.75cm}|>{\centering\arraybackslash}m{2.75cm}|}
\hline
$\alpha_S(M_Z^2)$ & $\Delta \chi^2_{\rm global}$ (NNLO) & $\Delta \chi^2_{\rm global}$ (aN$^{3}$LO) & $\Delta \chi^2_{\rm global}$ (NNLO dijets) & $\Delta \chi^2_{\rm global}$ (aN$^{3}$LO dijets) \\ \hline
0.112 & 293.0 & 190.6 & 388.3 & 185.3 \\ 
0.113 & 164.3 & 125.7 & 277.6 & 116.8 \\ 
0.114 & 109.9 & 70.6 & 182.1 & 68.0 \\ 
0.115 & 54.2 & 32.5 & 106.0 & 35.1	\\ 
0.116 & 16.5 & 8.5 & 55.8 & 6.5 \\ 
0.117 & 1.0	& 0.0 & 19.4 & 0.0 \\ 
0.1171 & 0.0 & - & - & - \\
0.118 & 3.8 & 6.1 & 2.0 & 5.8 \\ 
0.1181 & - & - & 0.0 & - \\ 
0.119 & 29.4 & 27.4 & 4.9 & 25.7 \\ 
0.120 & 72.9 & 63.3 & 27.0 & 61.3 \\ 
0.121 & 140.5 & 117.0 & 68.5 & 110.5\\ 
0.122 & 223.6 & 164.9 & 129.4 & 173.8\\ 
\hline
$N_{\rm pts}$ & 4534 & 4534 & 4157 & 4157 \\ \hline 
\end{tabular}
\vspace{0.1cm}
\caption {The fit quality of the global fits versus $\alpha_S(M_Z^2)$ at NNLO and aN$^{3}$LO for the default case including the inclusive jet data and for the case where this is replaced by dijet data and relative to their respective best fits. The number of datapoints is also given as this differs between the default and dijet cases.}
\label{tab:alphas_deltachisqs}
\end{table}
\end{center}

\subsection{Individual Dataset $\as$ dependencies} \label{subsec:individualdatasets}

In addition to analysing the total global $\chi^2$ across all the datasets in the PDF fit, we can also consider the $\chi^2$ profiles of individual datasets within the context of the global fit as $\as$ is changed. These provide information about the values of $\as$ preferred by different datasets\footnote{Note these are simply the $\chi^2$ contributions of each dataset to the totals shown in Fig.~\ref{fig:chi2_profiles_total_NNLOaN3LO}, i.e. the change in $\chi^2$ of this dataset with $\as$ as the whole PDF is refit.}.

\begin{figure}
\begin{center}
\includegraphics[scale=0.225]{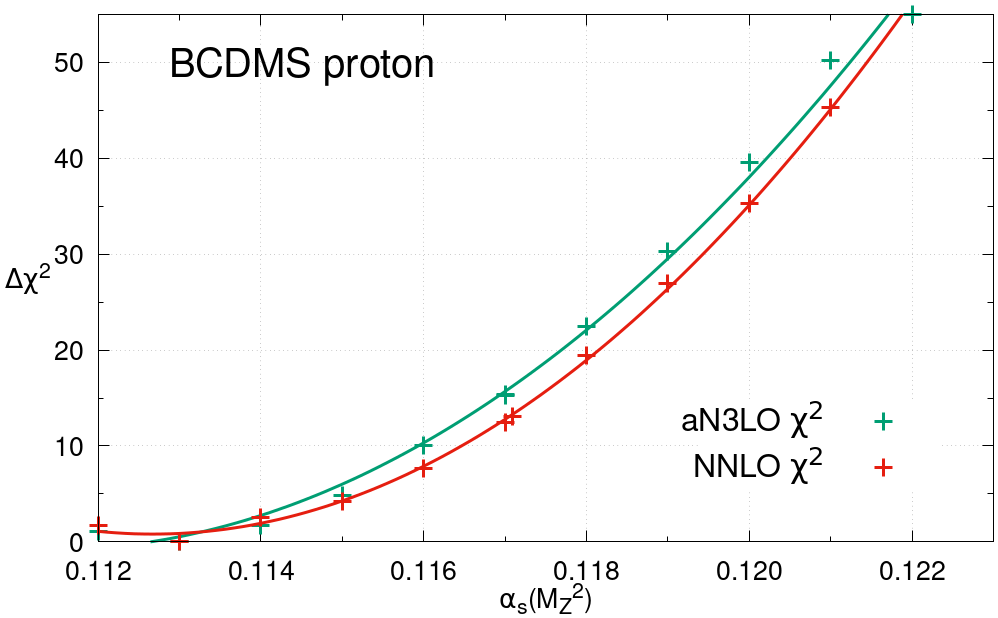}
\includegraphics[scale=0.225]{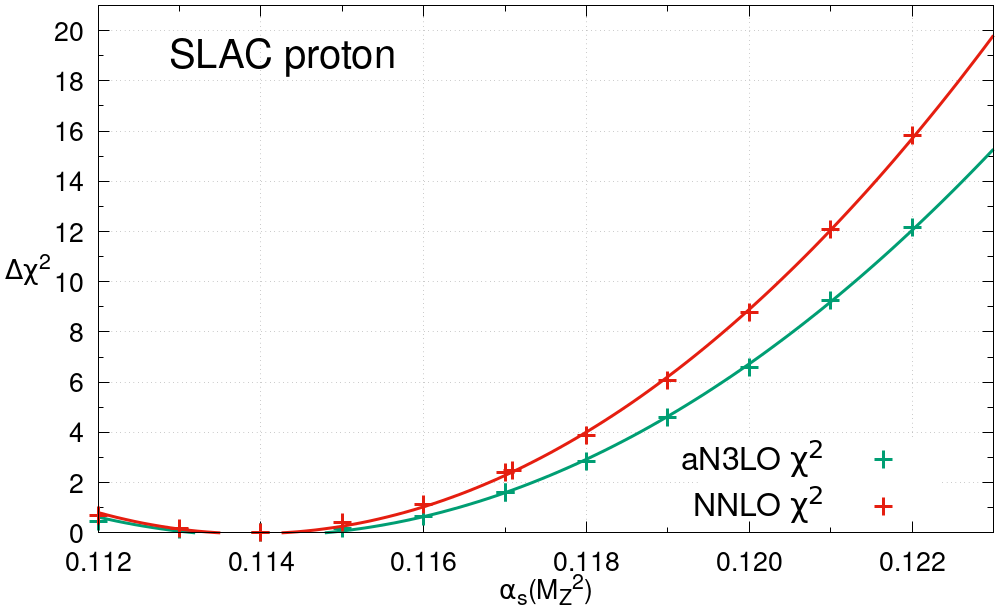}
\includegraphics[scale=0.225]{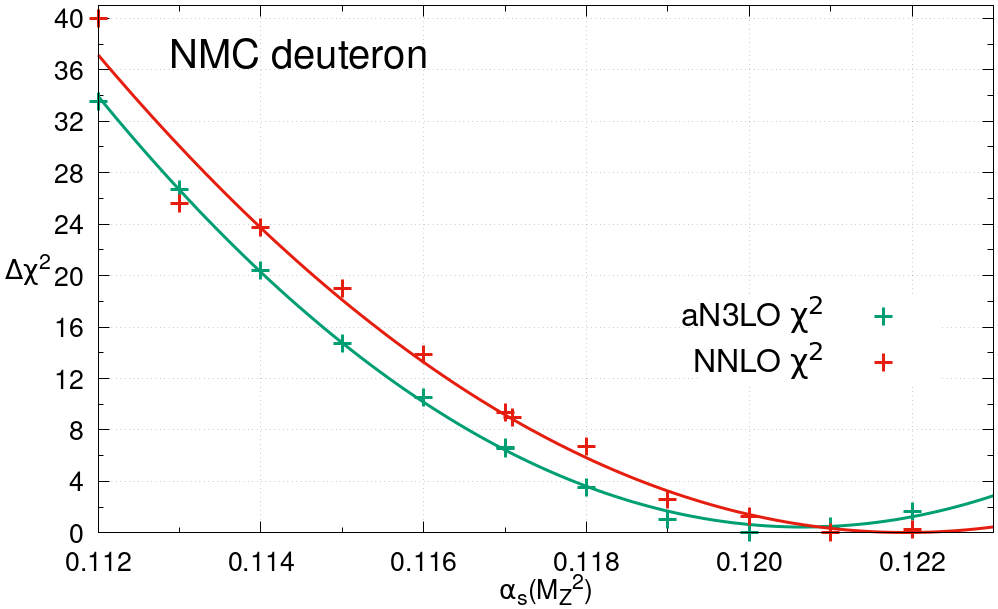}
\includegraphics[scale=0.225]{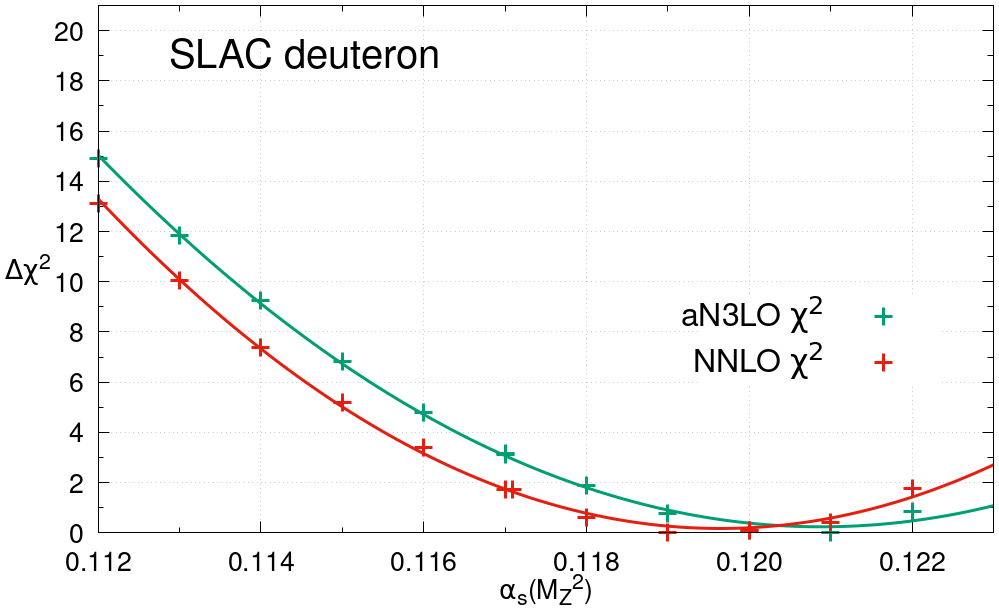}
\caption{\sf The individual datasets $\Delta \chi^2 = \chi^2 - \chi^2_0$ for different values of $\as$, within the global PDF fits at NNLO (red) and aN$^{3}$LO (green). This figure provides the profiles with $\as$ for a small selection of the fixed target deep inelastic scattering datasets included.}
\label{fig:individual_fixedtargerDIS_chisqs}
\end{center}
\end{figure}

For brevity we will show only the $\chi^2$ profiles of a small selection of the datasets included in the global PDF fit here. We begin with a selection of the most sensitive fixed target deep inelastic scattering (DIS) experiments from BCDMS \cite{Benvenuti:1989rh}, NMC \cite{Arneodo:1996qe} and SLAC \cite{Whitlow:1991uw,Whitlow:1990gk} in Fig.~\ref{fig:individual_fixedtargerDIS_chisqs}. As observed in \cite{Cridge:2021qfd}, the $F_2^p$ data favour values of $\as$ substantially lower than the global best fit, of around 0.113 and 0.114 respectively at both NNLO and aN$^{3}$LO. The DIS coefficient functions are very largely known at N$^{3}$LO \cite{Vermaseren:2005qc} and implemented in our aN$^{3}$LO PDF fit \cite{McGowan:2022nag} so these data are analysed to higher order here than in previous global PDF $\as$ determinations. The fixed target data are mostly sensitive to high $x$, and provide  relatively clean measurements of $\as$, as they depend largely on non-singlet PDF combinations for which the N$^3$LO splitting functions are known with less uncertainty than the singlet, particularly at large $x$. This contrasts with HERA data, which generally being at higher $Q^2$ and lower $x$ are more sensitive to the singlet and hence correlations between $\as$ and the gluon PDF, which reduces their sensitivity. The sensitivity to $\as$ in the BCDMS, SLAC and other fixed target $F_2^p$ data comes largely through the DGLAP evolution across scales between these data and higher $Q^2$ data and also within the data in the case of BCDMS, where the data cover a significant range in $Q^2$. In the latter case the reduction in $\as$ acts to reduce the fall in the structure function with $Q^2$, which is preferred by the BCDMS data \cite{Martin:1998sq}. Considering instead deuteron fixed target DIS, $F_2^d$ at NMC and SLAC we observe the opposite trend, with larger values of $\as$ preferred of around 0.120, indicating some tension with the BCDMS and SLAC $F_2^p$ data. Deuteron corrections are included for these datasets as described in \cite{Bailey:2020ooq}. In all 4 cases shown (and others not shown) there is good consistency between NNLO and aN$^{3}$LO $\as$ $\chi^2$ profiles for each dataset and hence in their preferred $\as$ values.

\begin{figure}
\begin{center}
\includegraphics[scale=0.225]{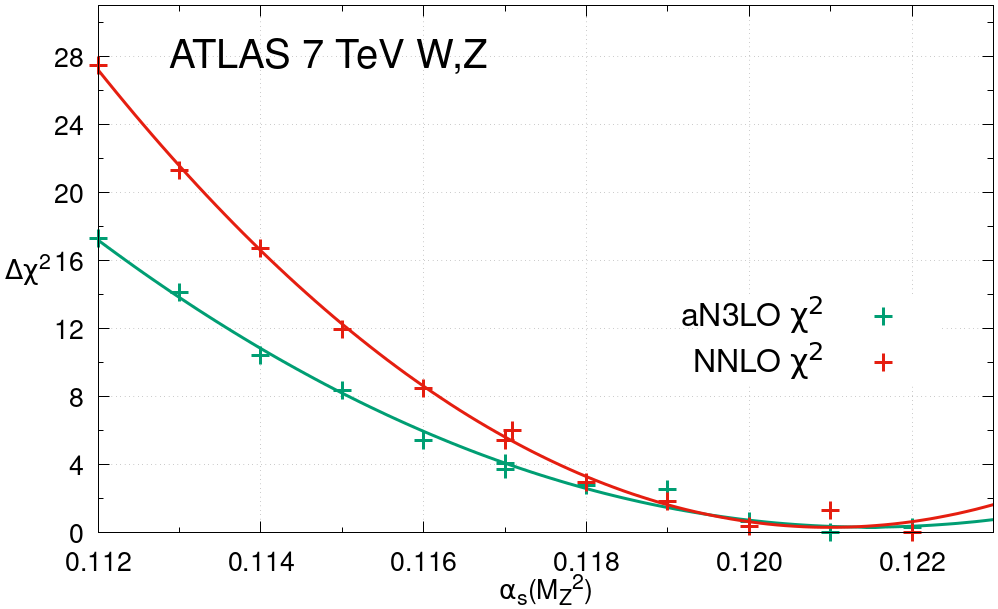}
\includegraphics[scale=0.225]{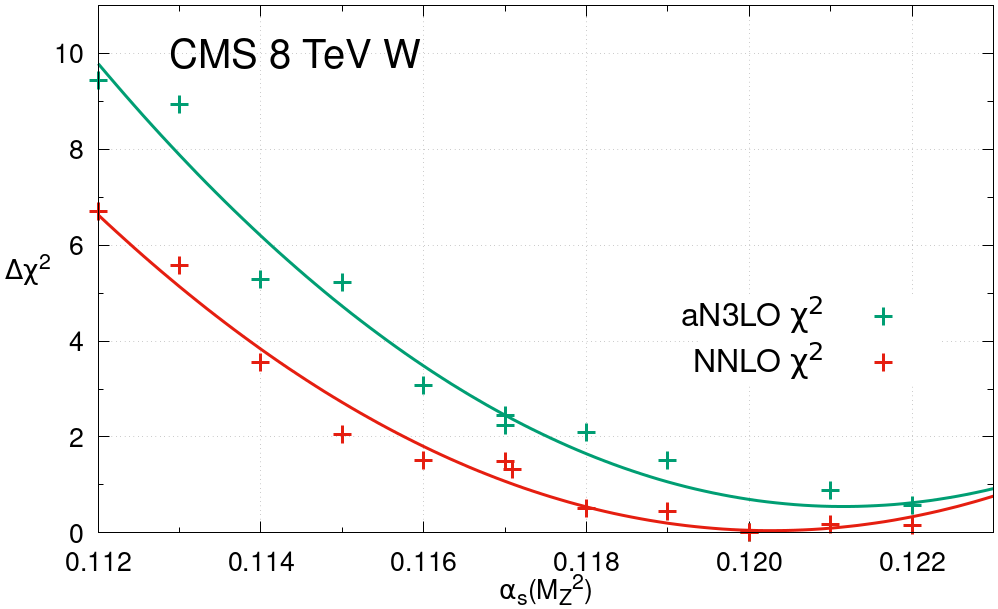}
\includegraphics[scale=0.225]{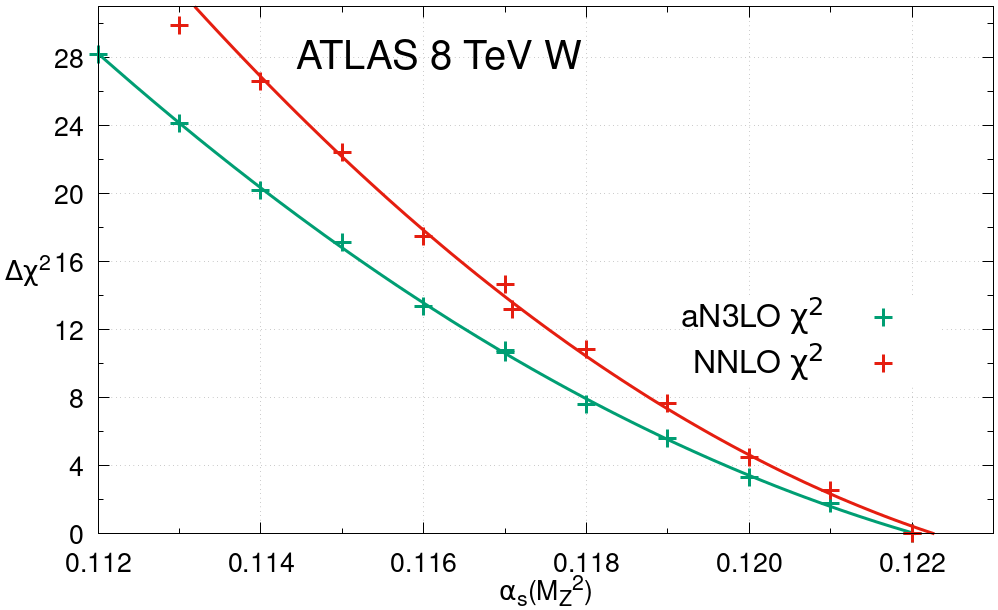}
\includegraphics[scale=0.225]{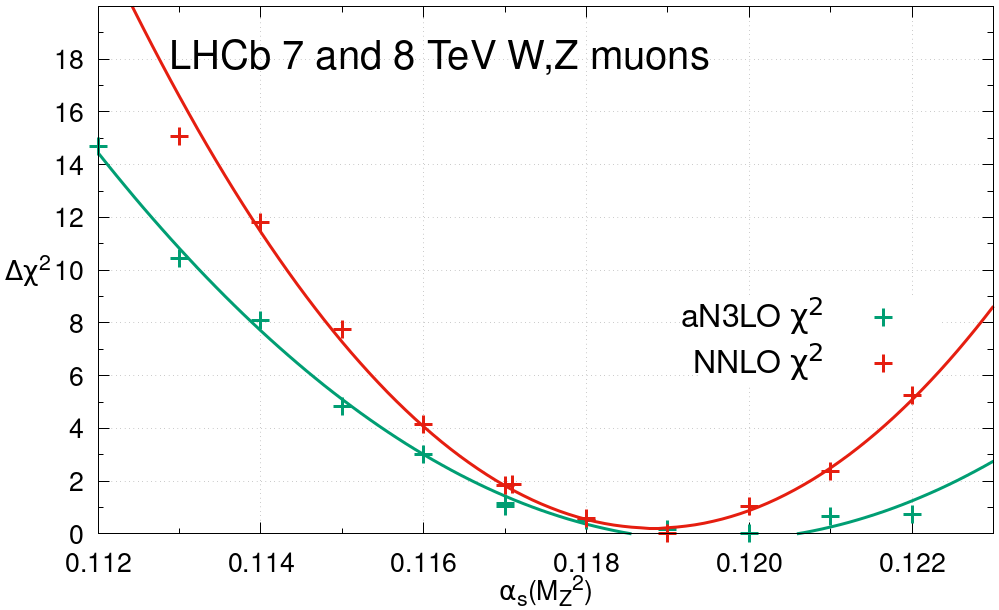}
\caption{\sf The individual datasets $\Delta \chi^2 = \chi^2 - \chi^2_0$ for different values of $\as$, within the global PDF fits at NNLO (red) and aN$^{3}$LO (green). This figure provides the profiles with $\as$ for a small selection of the collider Drell-Yan datasets included.}
\label{fig:individual_DY_chisqs}
\end{center}
\end{figure}

In addition to DIS datasets providing $\as$ sensitivity in the PDF fit, more recent collider data from the LHC and elsewhere provide further complementary sensitivity. We begin by providing the $\chi^2$ profiles for a variety of LHC Drell-Yan datasets in Fig.~\ref{fig:individual_DY_chisqs}. We show a selection of ATLAS\cite{ATLASWZ7f,ATLASW8}, CMS\cite{CMSW8} and LHCb\cite{LHCbZ7,LHCbWZ8} data at 7 and 8~TeV for illustration. Whilst these data have limited direct sensitivity to $\as$ through the $\as$ dependence of their cross sections, their precision provides notable $\as$ dependence in the context of the global PDF fit due to the impact of $\as$ on the PDF themselves. These datasets consistently indicate a preference for an $\as$ value larger than the best fit, of 0.119 and higher, as noted in \cite{Cridge:2021qfd}. Again we observe consistency between the NNLO and aN$^{3}$LO $\as$ profiles, though now the aN$^{3}$LO profiles are usually somewhat shallower, indicating a slightly reduced sensitivity to $\as$. This is a reflection of the unknown N$^{3}$LO K-factors for these processes\footnote{Whilst there was been recent progress in the determination of cross-sections for several processes at N$^{3}$LO\cite{Caola:2022ayt}, including neutral and charged current Drell-Yan \cite{Duhr:2020seh,Duhr:2020sdp,Gehrmann:DYN3LO,duhr:DY2021} both total and differential in rapidity, these are not yet provided in a form for utilisation in PDF fits, in particular differential over all the required variables and with fiducial cuts applied.}, which contribute additional MHOU uncertainties to the PDFs and in turn the inclusion of these theoretical uncertainties reduce the $\as$ sensitivity mildly.

\begin{figure}
\begin{center}
\includegraphics[scale=0.225]{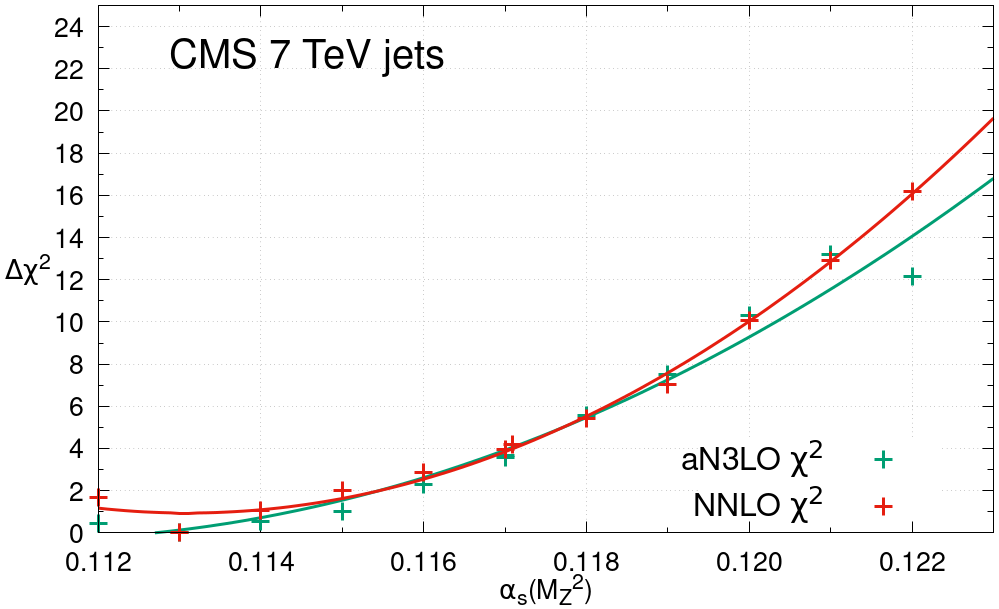}
\includegraphics[scale=0.225]{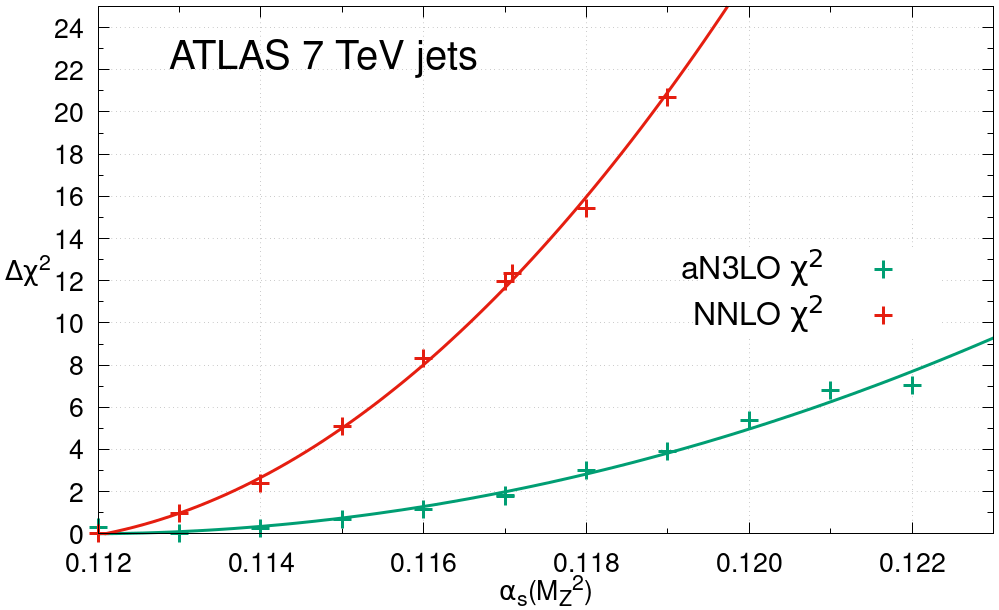}
\includegraphics[scale=0.225]{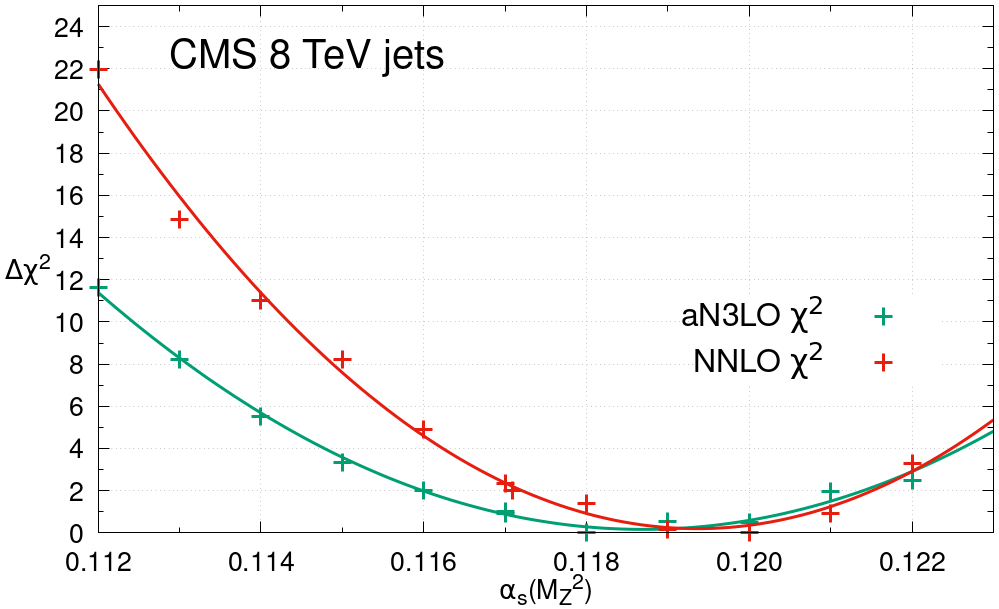}
\includegraphics[scale=0.225]{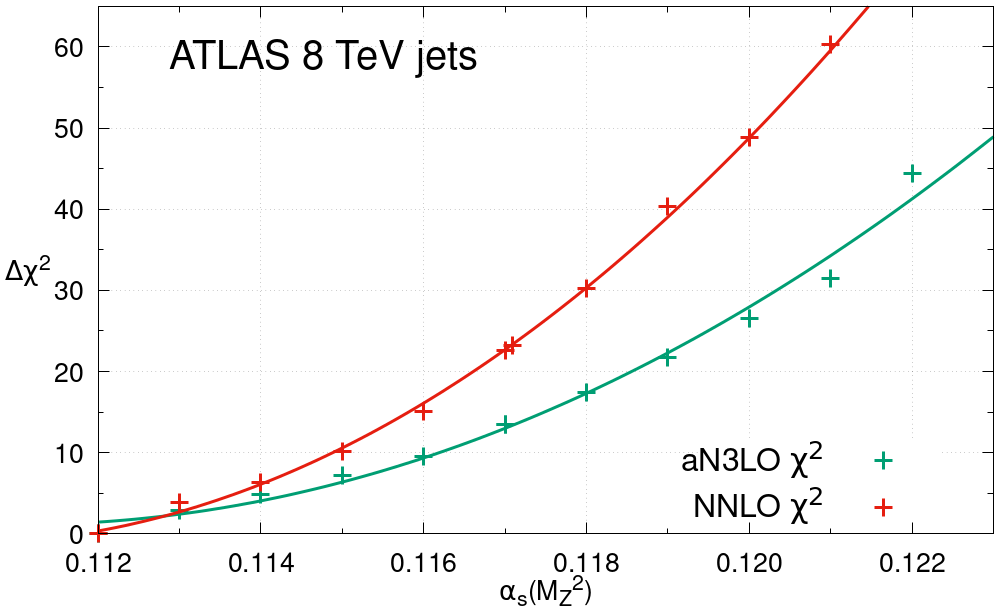}
\caption{\sf The individual datasets $\Delta \chi^2 = \chi^2 - \chi^2_0$ for different values of $\as$, within the global PDF fits at NNLO (red) and aN$^{3}$LO (green). This figure provides the profiles with $\as$ for a small selection of the collider inclusive jet datasets included.}
\label{fig:individual_jet_chisqs}
\end{center}
\end{figure}

In contrast to Drell-Yan data, inclusive jet data have significant direct $\as$ sensitivity through the cross-section. Fig.~\ref{fig:individual_jet_chisqs} illustrates the $\chi^2$ profiles for several of these datasets with $\as$. As expected, notable $\as$ sensitivity is observed with the CMS \cite{CMS:2014nvq} and ATLAS 7~TeV inclusive jets \cite{ATLAS:2014riz} favouring lower values of $\as$, around 0.112 at both NNLO and aN$^{3}$LO. The ATLAS 8~TeV inclusive jet data \cite{ATLAS:2017kux} favour a similarly low value of $\as$, though in contrast the CMS 8~TeV inclusive jet data \cite{CMS:2016lna} show some tension with these, preferring $\as$ around 0.119 at both NNLO and aN$^{3}$LO. These results are consistent with the pulls seen previously of these datasets on the high $x$ gluon \cite{Bailey:2020ooq,Jing:2023isu,Cridge:2021qjj,PDF4LHCWorkingGroup:2022cjn}, with those which prefer a larger high $x$ gluon generally favouring a smaller value of $\as$. This is as expected given the correlation between the high $x$ gluon PDF and $\as$ at large $x$, as indicated in Fig.~\ref{fig:aN3LO_PDFalphas_corrs}. In addition we note that, as for the Drell-Yan datasets, the aN$^{3}$LO profiles are often notably shallower than at NNLO due to the inclusion of a theoretical uncertainty from the missing N$^{3}$LO cross-sections for these data. A more detailed analysis of these data follows in Section~\ref{sec:dijets}, where the impacts of fitting the inclusive jet or dijet data on $\as$ are examined.

\begin{figure}
\begin{center}
\includegraphics[scale=0.225]{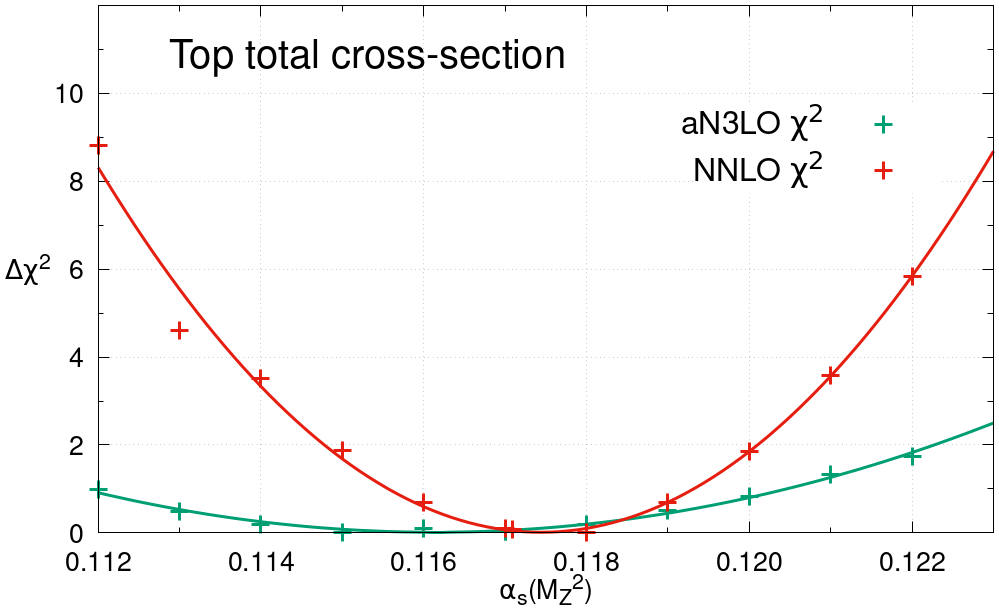}
\includegraphics[scale=0.225]{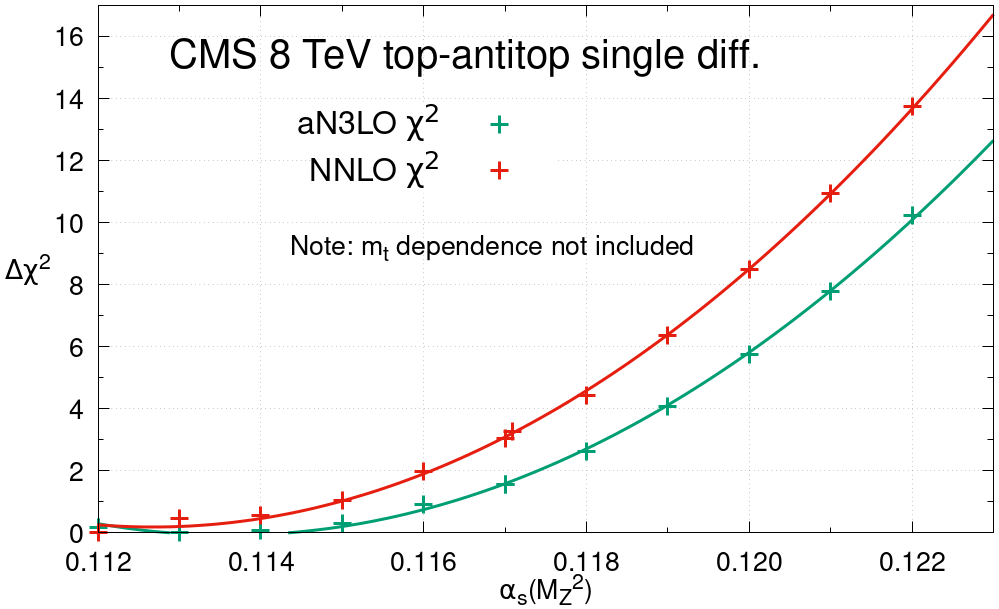}
\includegraphics[scale=0.225]{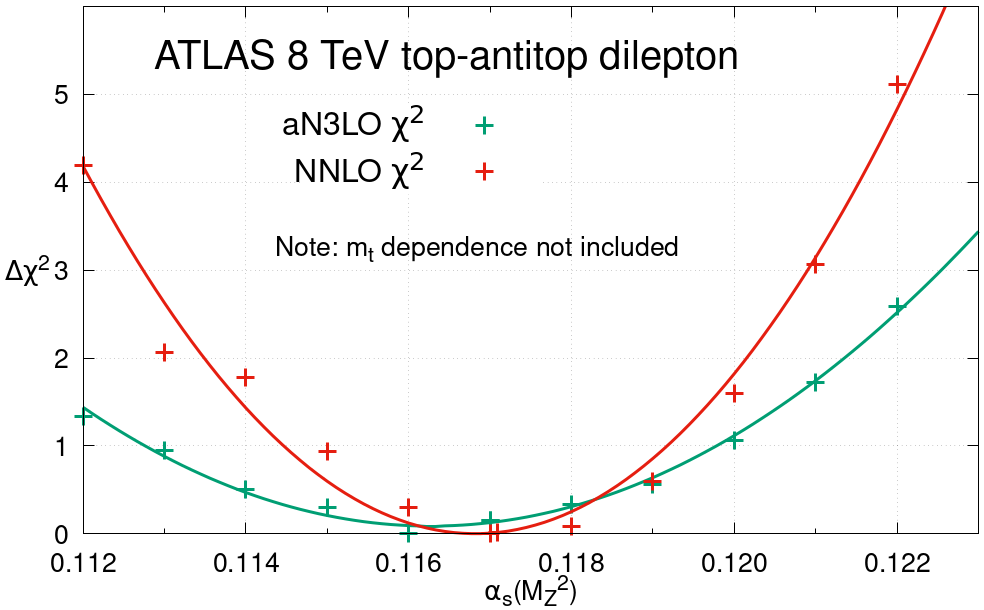}
\includegraphics[scale=0.225]{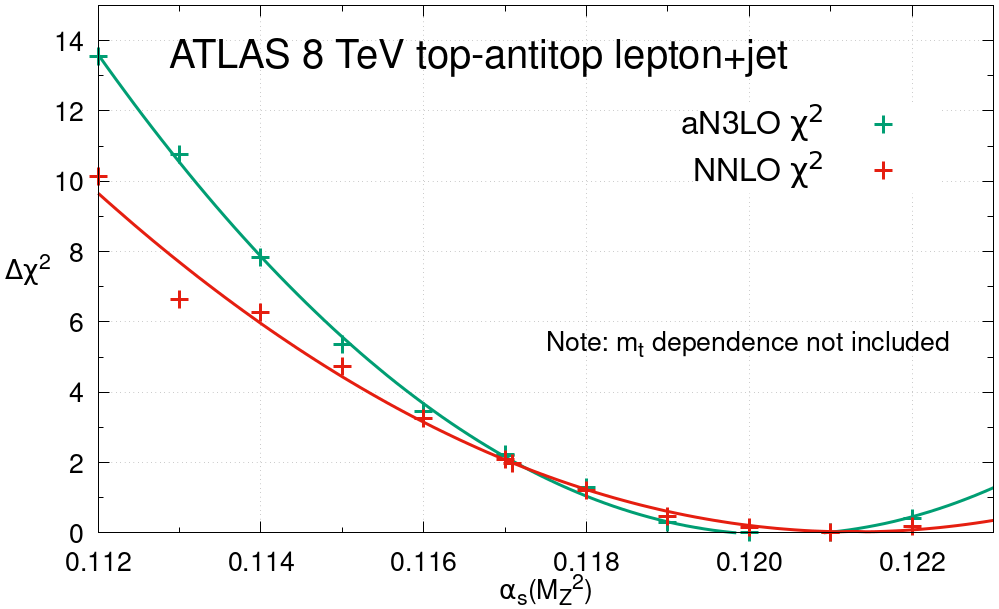}
\caption{\sf The individual datasets $\Delta \chi^2 = \chi^2 - \chi^2_0$ for different values of $\as$, within the global PDF fits at NNLO (red) and aN$^{3}$LO (green). This figure provides the profiles with $\as$ for a small selection of the collider top production datasets included. }
\label{fig:individual_top_chisqs}
\end{center}
\end{figure}

Next we discuss the $\as$ sensitivity of top quark pair production data. As for inclusive jet data, these are also expected to show significant sensitivity to $\as$, being $\mathcal{O}(\alpha_S^2)$ at leading order. In Fig.~\ref{fig:individual_top_chisqs}, we present the changes in $\chi^2$ as $\as$ is scanned from 0.112 to 0.122 for four datasets (a subset of the top quark datasets included in the global fit). These are: the top total cross-section data from the Tevatron, ATLAS and CMS \cite{Tevatron-top,ATLAS-top7-1,ATLAS-top7-2,ATLAS-top7-3,ATLAS-top7-4,ATLAS-top7-5,ATLAS-top7-6,CMS-top7-1,CMS-top7-2,CMS-top7-3,CMS-top7-4,CMS-top7-5,CMS-top8}; CMS 8~TeV top quark pair production differential in top-antitop pair rapidity in the lepton+jets channel \cite{CMSttbar08_ytt}; ATLAS 8~TeV data in the same lepton+jets channel but multi-differential in the top-quark pair invariant mass, top-quark/antiquark transverse momentum and the individual and pairwise rapidities \cite{ATLASsdtop}; and finally the ATLAS 8~TeV top-antitop production in the dilepton channel single differential in the top pair rapidity \cite{ATLASttbarDilep08_ytt}. The pulls on $\as$ are consistent between NNLO and aN$^{3}$LO, and with our previous NNLO study \cite{Cridge:2021qfd}. The CMS 8~TeV $t\bar{t}$ single differential data favour a lower $\as$, in contrast to the preference for a higher $\as$ observed in the ATLAS 8~TeV $t\bar{t}$ multi-differential data in the same lepton+jets channel. The top total cross-section data and ATLAS 8~TeV dilepton data both constrain $\as$ to be close to the best fit at NNLO, as observed previously, while at aN$^{3}$LO they favour slightly lower $\as \sim 0.116$. The main difference at aN$^{3}$LO with these latter two datasets is again the shallower nature of the $\chi^2$ profiles, placing less tight bounds on $\as$ due to the inclusion of a theoretical uncertainty for the MHOUs in the N$^{3}$LO cross-section. The results shown here are all shown at a fixed value of the top mass, however it was shown at NNLO within MSHT \cite{Cridge:2023ztj} that, at least in the case of the total top cross-section and the lepton+jet channels, that the dependence of the best fit on $\as$ and $m_t$ is relatively independent in the neighbourhood of the best fit. Indeed our results here at fixed $m_t$ are consistent with the $\as$ bounds investigated in \cite{Cridge:2023ztj}.

\begin{figure}
\begin{center}
\includegraphics[scale=0.4]{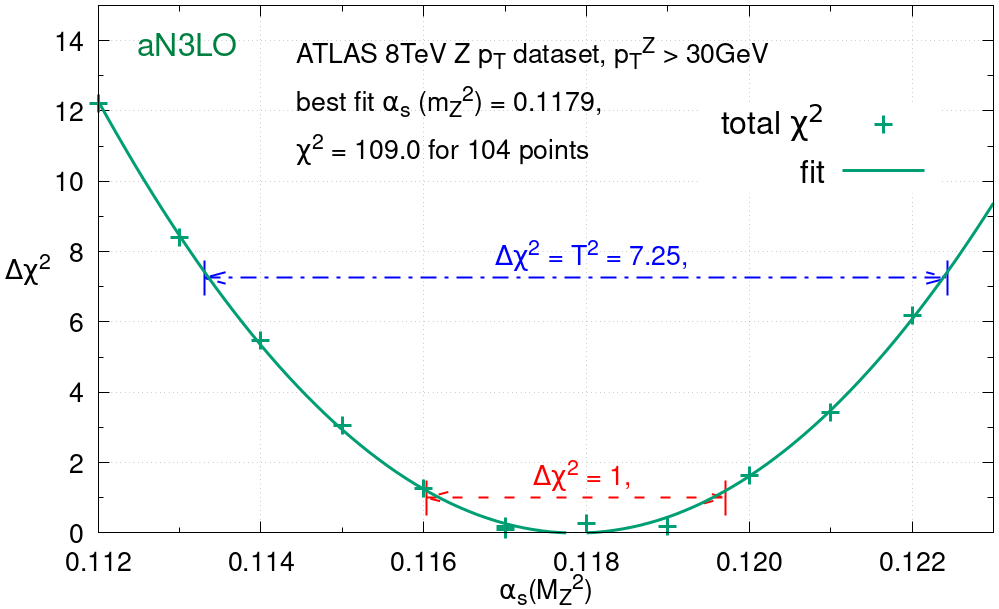}
\caption{\sf The $\Delta \chi^2 = \chi^2 - \chi^2_0$ for different values of $\as$, within the global PDF fit at aN$^{3}$LO for the ATLAS 8~TeV $Z$ $p_T$ dataset. In addition the $\Delta\chi^2=1,T^2$ limits are marked, the latter correspond to the bounds these data provide on $\alpha_S(M_Z^2)$ in the MSHT global fit approach with a dynamic tolerance.}
\label{fig:individual_ZpT_chisqs}
\end{center}
\end{figure}

Finally, a dataset which has been a focus of attention for its $\as$ sensitivity is the ATLAS 8~TeV $Z$ $p_T$ data\cite{ATLASZpT}. A recent measurement of these data \cite{atlascollaboration2023precise} was utilised to extract $\as$, with apparently significant $\as$ sensitivity in the low $p_T^Z$ region around the Sudakov peak\cite{ATLAS:2023lhg}. As these data are included at fixed order in global PDF fits we utilise a cut of $p_T^Z>  30\,{\rm GeV}$ to restrict ourselves to the region where this is valid. It was illustrated in \cite{McGowan:2022nag} that these data show a significant improvement in fit quality at aN$^{3}$LO relative to NNLO, which was subsequently analysed in more detail in \cite{Cridge:2023ryv} and concluded to be a sign of the necessity of the inclusion of higher order effects in the PDFs to fit these data. This also studied the impact of raising the $p_T^Z$ cut above $30\, {\rm GeV}$ and noted the same trend. In any case, given the precision of these data, even in the absence of the low $p_T^Z$ region some $\as$ sensitivity remains and we can analyse this within the context of the MSHT20 global PDF fit at aN$^{3}$LO. The $\Delta\chi^2$ profile at aN$^{3}$LO as $\as$ is changed is shown in Fig.~\ref{fig:individual_ZpT_chisqs}, and we can see that these data prefer $\as \approx 0.118$. We may also utilise the changes in $\chi^2$ with $\as$ to place bounds on $\as$, as described in more detail in Section~\ref{subsec:bounds}.

\subsection{Uncertainty and Bounds on $\alpha_S(M_Z^2)$} \label{subsec:bounds}

The $\chi^2$ profiles as $\as$ is changed also allow bounds to be set on $\as$ from each dataset within the context of the global PDF fit, and in turn to determine the overall uncertainty on our $\as$ determination. In order to do so, we utilise the same procedure as for the PDF eigenvectors to set bounds on the $\as$ eigenvector direction, which corresponds to our uncertainty on $\as$. This has been performed for the PDF uncertainties since MSTW08 \cite{Martin:2009iq} and was also utilised for the determination of bounds on the strong coupling in \cite{MMHTas,Cridge:2021qfd,Cridge:2023ztj}\footnote{In \cite{Cridge:2023ztj} this method is also used to set bounds on the top quark pole mass in the MSHT20 global PDF fit.}. 

The method used is known as the dynamical tolerance, and an outline of this is as follows. First the 68\% confidence level $\Delta\chi^2$ departure for each dataset from its value at the global best fit is determined by rescaling it by the 68\% confidence interval of a $\chi^2$ distribution with $N_{\rm pts}$. Then, as is done for each PDF eigenvector direction, we analyse the changes in $\chi^2$ for each individual dataset as $\as$ is scanned away from the global minimum and determine its bounds on $\as$ when it exceeds the previously determined 68\% confidence level $\Delta\chi^2$ for that dataset. Repeating this analysis for all datasets we build up a set of upper and lower bounds on $\as$, the most stringent of which become our global bounds, setting the uncertainty on our $\as$ extraction. In this way each eigenvector direction, or in this case scanning $\as$ upwards and downwards, is regarded as an alternative hypotheses and each dataset is regarded as providing a bound on it once its  $\Delta\chi^2$ exceeds $\chi_{n,0}^2 (\frac{\xi_{68}}{\xi_{50}}-1)$, where $\chi_{n,0}^2$ is its value at the global minimum and $\xi_{i}$ is the ith percentile of the $\chi^2$ distribution. Thus $\xi_{50}\approx N$ and for small $N$ $\xi_{68}\approx N+\sqrt{2N}$. This is a weaker criterion than the textbook parameter fitting definition and so results in larger uncertainties \cite{collins2001tests}. The resulting larger $\Delta\chi^2=T^2$ criterion for each dataset is referred to as the tolerance, $T$. The justification of this enlarged $\Delta\chi^2$ tolerance criterion has been examined previously \cite{collins2001tests,Martin:2009iq,Harland_Lang_2015}, and accounts for tensions among datasets as well as experimental, methodological and theoretical issues which prevent the application of the textbook $\Delta\chi^2=1$ criterion for the global dataset. 

Bounds of individual datasets on $\as$ within the MSHT global PDF fits are thus set by determining when the $\Delta\chi^2$ profiles in Figs.~\ref{fig:individual_fixedtargerDIS_chisqs}-\ref{fig:individual_ZpT_chisqs} cross their appropriate limits. This is explicitly shown in the context of the ATLAS 8~TeV $Z$ $p_T$ dataset\cite{ATLASZpT} in Fig.~\ref{fig:individual_ZpT_chisqs}, where the above procedure determines that its 68\% confidence limit corresponds to $\Delta\chi^2=T^2=7.25$. As can be seen, the resulting bounds from this data in the context of the MSHT20 global fit are relatively weak (as explained earlier the cut on $p_T^Z$ limits its $\as$ sensitivity) with a lower bound of $\approx 0.113$ and an upper bound of $\approx 0.122$ set. On the other hand, in the context of individual dataset $\as$ determinations, outside of the global PDF fits, whilst one must correctly account for the PDF eigenvector tolerances \cite{Schmidt_2018,Hou_2021} in profiling of the uncertainties on the PDFs, the $\Delta\chi^2$ criterion applied on the extraction of the parameter of interest is then up to those performing the study, with $\Delta\chi^2=1$ often being used for $\as$\footnote{Such studies also have the additional caveat of the potential issues of correlations of PDFs and $\as$ being neglected \cite{Forte:2020pyp}.}. This will not replicate the results found within the global fit, where the tolerance is also used for $\as$. The difference of applying a $\Delta\chi^2=1$ bound\footnote{Though still within the context of the global fit as before and refitting at each $\as$.} can be seen in Fig.~\ref{fig:individual_ZpT_chisqs},  and results in tighter bounds, though still weaker than the overall bounds found across the whole global fit dataset.

\begin{figure}
\begin{center}
\includegraphics[scale=0.41]{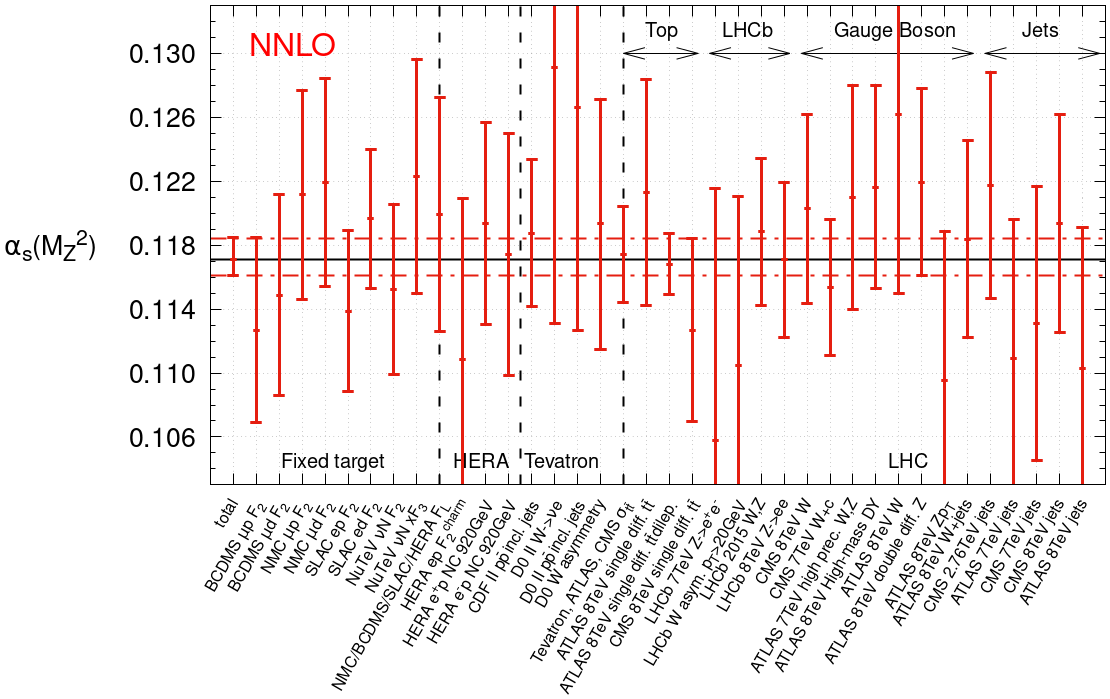}
\includegraphics[scale=0.41]{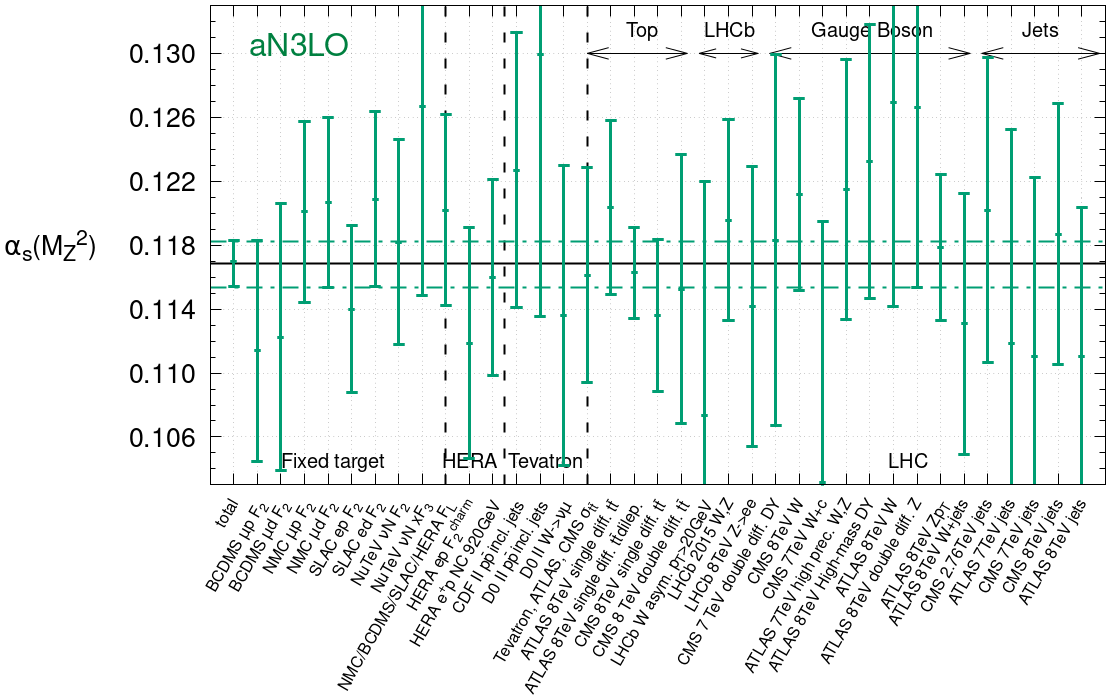}
\caption{\sf The overall and dataset by dataset best fit value and upper and lower bounds on $\alpha_S(M_Z^2)$, for a selection of the datasets in the global fit. The overall upper and lower bounds are given by the horizontal dashed lines, whilst the coloured vertical solid lines show the individual dataset bounds. The upper plot is the NNLO fit and the lower is the aN$^{3}$LO fit.}
\label{fig:alphasbounds_NNLOaN3LO}
\end{center}
\end{figure}

Repeating this analysis across all of the datasets in the PDF fit, at both NNLO and aN$^{3}$LO, we obtain the results shown in Fig.~\ref{fig:alphasbounds_NNLOaN3LO}. Beginning with the results at NNLO in the upper plot of Fig.~\ref{fig:alphasbounds_NNLOaN3LO}, the results are consistent with those determined previously at this order in \cite{Cridge:2021qfd}. We observe the tightest upper bound on $\as$ comes from the BCDMS $F_2$ proton data\cite{Benvenuti:1989rh}, which provide the bound $\Delta\as = +0.0014$ in the upwards direction. This results from the behaviour seen in Fig.~\ref{fig:individual_fixedtargerDIS_chisqs} (upper left) and the preference for the data to slow the fall of the structure function with $Q^2$. For comparison a very similar upper bound of $\Delta\as = +0.0012$ was observed in our previous study. The SLAC $F_2$ proton data \cite{Whitlow:1990gk,Whitlow:1991uw} and ATLAS 8~TeV $Z$ $p_T$ \cite{ATLASZpT} provide slightly weaker upper bounds of $\Delta\as = +0.0018$, though the latter is quite poorly fit at NNLO and there is evidence for the need for aN$^{3}$LO to describe these precise data well \cite{McGowan:2022nag,Cridge:2023ozx}. The former also provided a very similar bound in \cite{Cridge:2021qfd}. Several of the inclusive jet datasets are also able to place (weaker) upper bounds on $\as$, with the most stringent being the ATLAS 8~TeV jets\cite{ATLAS:2017kux} for which $\Delta\as = +0.0020$. This fits with the general observation made in Section~\ref{subsec:individualdatasets} of DIS and inclusive jet datasets often favouring lower values of $\as$ in the global fit and thus providing upper bounds, also observed in \cite{Cridge:2021qfd}. Meanwhile, for the lower bounds on $\as$ the tightest bound at NNLO is given by the ATLAS 8~TeV $Z$ data\cite{ATLAS8Z3D} for which $\Delta\as=-0.0010$, this provided a similar bound in our previous study \cite{Cridge:2021qfd}. The next strongest bound comes from the NMC $F_2$ deuteron data\cite{Arneodo:1996qe} for which  $\Delta\as=-0.0017$, whilst the ATLAS 8~TeV High Mass Drell-Yan data~\cite{ATLASHMDY8} and SLAC $F_2$ deuteron data~\cite{Whitlow:1990gk,Whitlow:1991uw} provide $\Delta\as=-0.0018$. These are again consistent with the picture observed in Section~\ref{subsec:individualdatasets} and previously in \cite{Cridge:2021qfd}, where Drell-Yan and deuteron datasets tend to favour larger values of $\as$ within the global PDF fit and therefore place lower bounds on $\as$. The general observation of Drell-Yan data favouring larger values of $\as$ and some DIS datasets (in particular the BCDMS proton data) preferring smaller values of $\as$ was also noted in \cite{Hou:2019efy}, and similarly in \cite{Ball_2018} with slight differences. Finally, Fig.\ref{fig:alphasbounds_NNLOaN3LO} also emphasises the robustness of the $\as$ uncertainty determination from global PDF fits, with several datasets able to provide upper and lower bounds, it is not reliant on any one dataset.

Considering now the aN$^{3}$LO PDF fit, we can for the first time obtain bounds on $\as$ in an aN$^{3}$LO PDF fit. Repeating this analysis on the datasets within the global PDF fit at aN$^{3}$LO we obtain the bounds shown in Fig.~\ref{fig:alphasbounds_NNLOaN3LO} (lower). Comparing this with the corresponding figure at NNLO, the similarities are immediately apparent. Once again the BCDMS $F_2$ proton data \cite{Benvenuti:1989rh} provides the tightest upper bound of $\Delta\as=+0.0013$. Now the next tightest bound is given by the charm structure function data $F_2^c$ from HERA\cite{H1+ZEUScharm}, for which $\Delta\as=+0.0020$. The charm structure function has greater sensitivity to $\as$ than light quarks as it is generated at $\mathcal{O}(\alpha_S)$. The ATLAS 8~TeV $Z$ $p_T$ data no longer provides a stringent upper bound at aN$^{3}$LO, as is clear from Fig.~\ref{fig:individual_ZpT_chisqs}, with it being considerably better fit at aN$^{3}$LO. Similarly, as seen in Fig.~\ref{fig:individual_jet_chisqs}, the sensitivity of the inclusive jet datasets to $\as$ (some of which provide relatively competitive upper bounds on $\as$ at NNLO) is reduced at aN$^{3}$LO due to the missing N$^{3}$LO K-factors and the associated included missing higher order uncertainty. In the downwards direction, the SLAC $F_2$ deuteron data  now provides the strongest bound\cite{Whitlow:1990gk,Whitlow:1991uw} closely followed by the NMC $F_2$ deuteron data \cite{Arneodo:1996qe} of $\Delta\as = -0.0016,-0.0017$ respectively. The Drell Yan datasets provide generally slightly weaker bounds at aN$^{3}$LO than NNLO, again reflecting the missing higher order uncertainty included, though the ATLAS 8~TeV $Z$ data \cite{ATLAS8Z3D} still provide a bound of $\Delta\as=-0.0017$. We therefore observe overall slightly weaker bounds on $\as$ at aN$^{3}$LO than at NNLO as a result of the missing higher order uncertainties incorporated into the PDF fit and their effects on the bounds of the LHC data. Nonetheless the overall bounds on $\as$ remain similar in size, as a result of the bounds from DIS data, which are now known theoretically almost completely at N$^{3}$LO. Again this emphasises the robustness of $\as$ determinations from global PDF+$\as$ fits, as whilst individual datasets may alter their $\as$ dependence, given several different datasets and different types of processes are combined to provide global bounds on $\as$ the net effect on the overall uncertainty determination is mild.

In summary, the overall determination of the best fit values of $\as$ and its uncertainty at NNLO and aN$^{3}$LO are:
\begin{equation*}
    \as({\rm NNLO}) = 0.1171 \pm 0.0014
\end{equation*}
\begin{equation*}
    \as({\rm aN^3LO}) = 0.1170 \pm 0.0016
\end{equation*}
Here we have taken the most conservative of the upper and lower bounds on $\as$ at each order and symmetrised for simplicity. The consistency of the determinations at NNLO and aN$^{3}$LO is clear, particularly considering the NLO determination in our previous study\cite{Cridge:2021qfd} of $\as({\rm NLO}) = 0.1203\pm0.0015$. In addition, the aN$^{3}$LO determination results in the slightly weaker bounds than at NNLO, very likely due to the inclusion of missing higher order theoretical uncertainties in the fit. These bounds on $\as$ correspond to a $\Delta\chi^2 = 13$ at NNLO and $\Delta\chi^2 = 16$ at aN$^{3}$LO. Both the NNLO and aN$^{3}$LO $\as$ determinations are consistent with the Particle Data Group (NNLO) world average of $0.1180\pm0.0009$\cite{Huston:2023ofk}.

\section{Examination of Approximate N3LO $\alpha_S(M_Z^2)$ sensitivity}

\subsection{Sensitivity of the Splitting functions} \label{subsec:sensitivity_splits}

At \anlo   the form of the splitting functions is allowed to vary in the fit, guided by a prior uncertainty band that is determined from the known information about these objects at the time of the release of the MSHT20\anlo set. We will in general  expect some dependence of the resulting splitting functions on the value of the strong coupling, and vice versa.

\begin{figure}
\begin{center}
\includegraphics[scale=0.24]{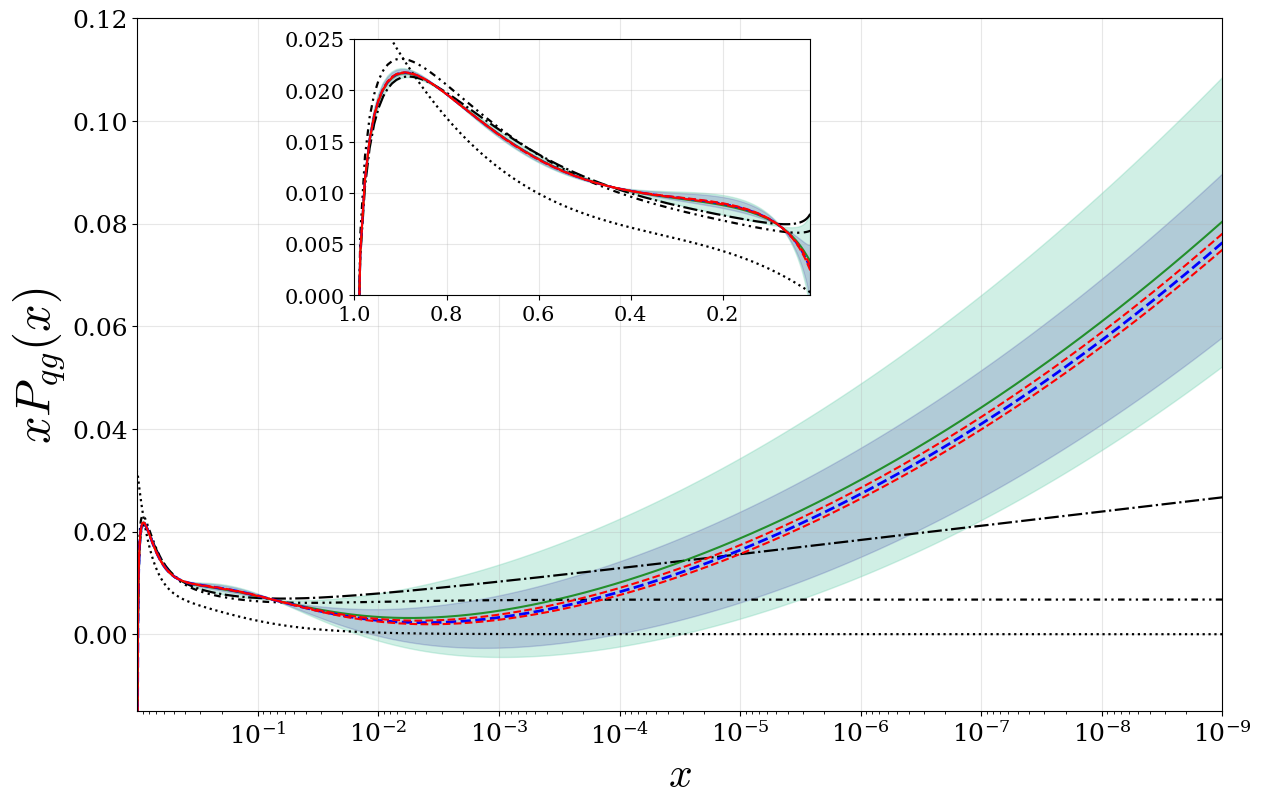}
\includegraphics[scale=0.24]{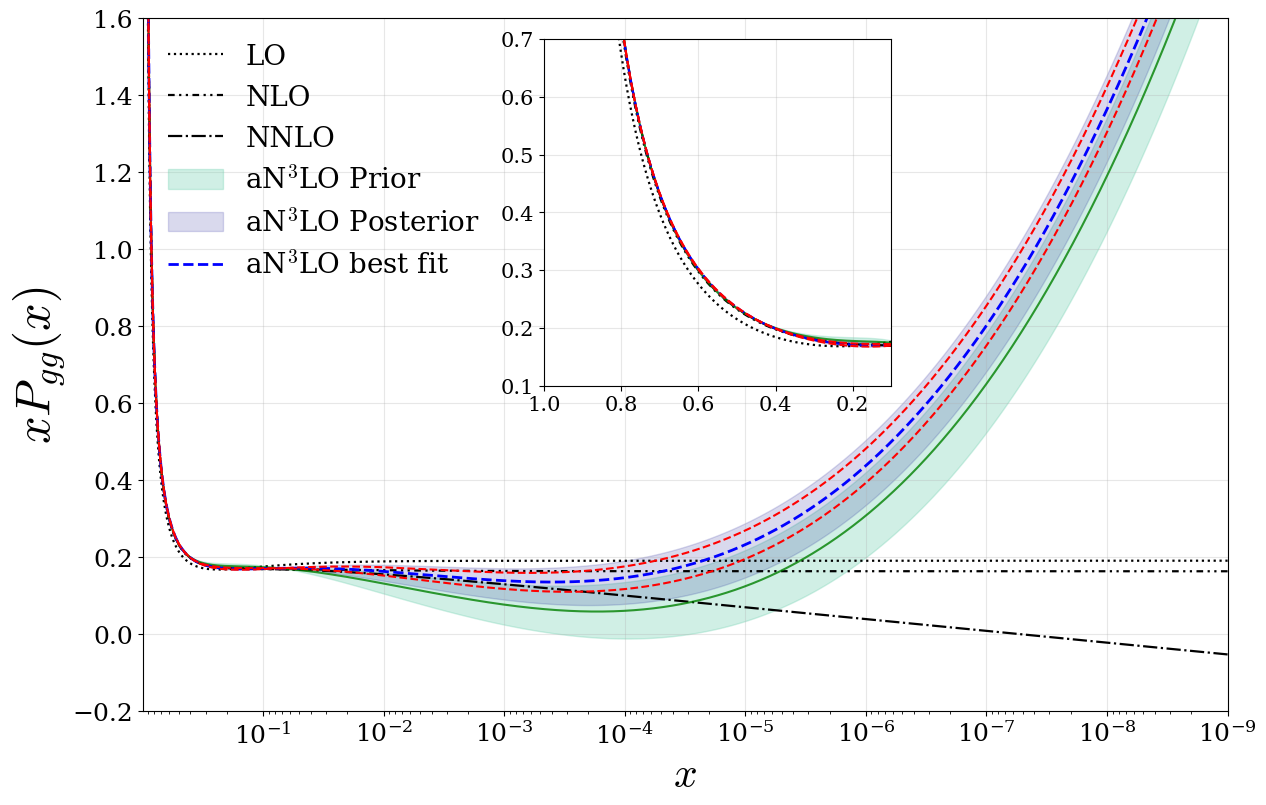}
\caption{\sf Posterior variations of the $qg$ and $gg$ splitting functions at the best fit value of $\alpha_S(M_Z^2)=0.117$, as well as for $\alpha_S(M_Z^2)=0.116$, 0.118 indicated by the dashed red curves. In the most visible part of the plots, i.e. below $x \approx 10^{-2}$ the lower (upper) curves correspond to the values of 0.116 (0.118). Also shown are the prior and lower order results, for comparison.}
\label{fig:split_vs_alphas_n3lo}
\end{center}
\end{figure}

It is therefore useful to examine the impact of the value of the strong coupling on the best fit splitting functions. This is shown in Fig.~\ref{fig:split_vs_alphas_n3lo} for the two cases that show the highest sensitivity; for other splitting functions the dependence is hardly visible on the plots. In particular, these show both the prior, and the posterior (at the best fit value of $\as=0.117$) uncertainty bands, as well as in the red dashed lines the best fit posterior splitting functions that result when $\as$ is varied by $\pm 0.001$. For demonstration purposes, we note that the splitting functions are shown at a fixed value of $\alpha_S=0.2$, which isolates the impact from the fit on the extracted splitting functions.

We can see that the largest dependence is for the gluon--gluon splitting function, which is as we might expect given the known correlation between the value of the strong coupling and the gluon PDF. For the larger value of $\as=0.118$, the splitting function is larger in the visible (lower $x$) region on the plots, while for the lower value of $\as=0.116$ it is lower. The size of the variation is nonetheless safely smaller than, although not negligible with respect to, the quoted posterior uncertainty, and in all cases these are within the original prior band. For the quark--gluon splitting function (and, as mentioned above the other cases not shown here) the dependence is much smaller. 

This therefore indicates only a mild sensitivity of the preferred splitting function on the value of $\as$ in our fit. Conversely, given this is a relatively small effect we can expect any dependence of the extracted value of $\as$ on the precise treatment of the splitting function uncertainties to be even smaller. Given additional information from more recent theoretical calculations of the splitting functions~\cite{Falcioni:2023luc,Falcioni:2023vqq,Falcioni:2023tzp,Moch:2023tdj,Gehrmann:2023cqm} is now available, this provides reassuring evidence that our analysis should not be significantly changed when this information is included in the PDF fit. Indeed, this is supported by the observation made earlier, that taking the updated splitting functions of \cite{Falcioni:2023luc,Falcioni:2023vqq,Falcioni:2023tzp,Moch:2023tdj} resulted in a best fit $\as$ very close to that we obtain in this work.

\subsection{Impact of Jet vs. Dijet production} \label{sec:dijets}

In~\cite{Cridge:2023ozx} we presented a detailed comparison of the impact of 7 and 8 TeV inclusive jet~\cite{CMS:2014nvq,CMS:2016lna,ATLAS:2014riz,ATLAS:2017kux} in comparison to dijet~\cite{ATLAS:2013jmu,CMS:2012ftr,CMS:2017jfq} data on the MSHT fit at up to \anlo order. In this section, we extend this analysis to examine the impact such data have on the extracted value of the strong coupling. Other than by allowing the value of $\as$ to vary, the baseline fits are identical to those presented in~\cite{Cridge:2023ozx}.

\begin{figure}
\begin{center}
\includegraphics[scale=0.22]{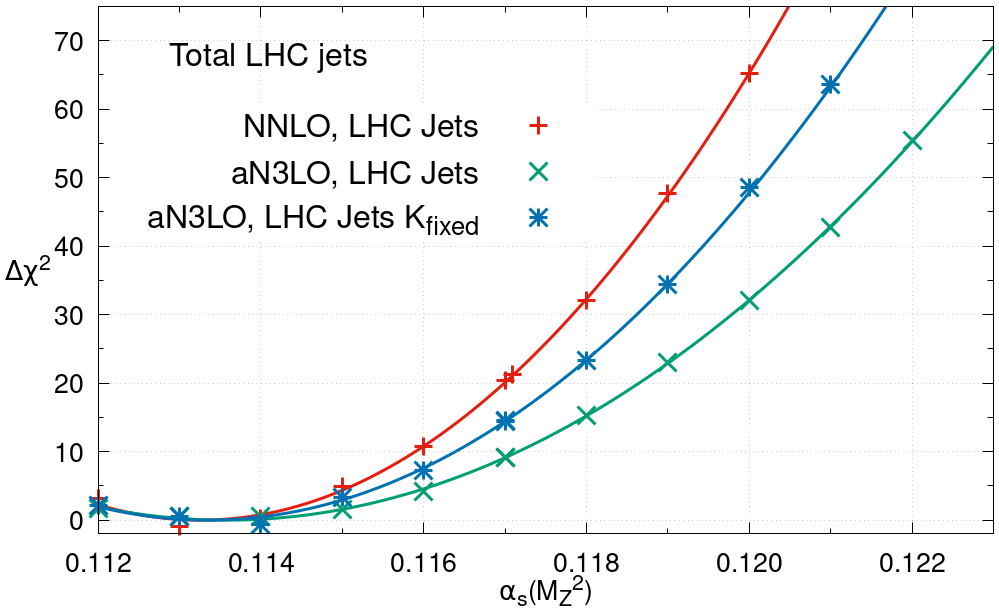}
\includegraphics[scale=0.22]{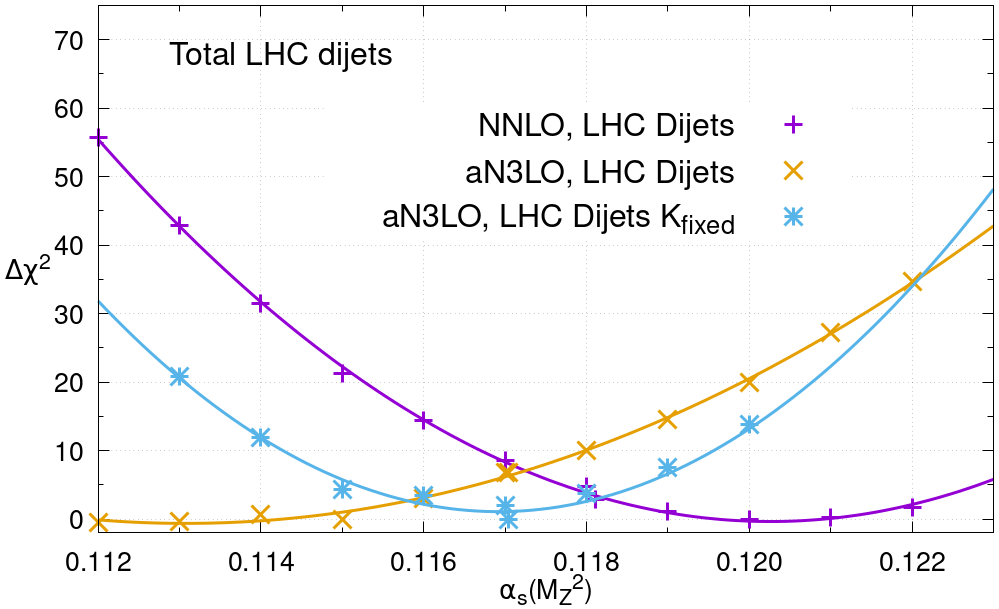}
\caption{\sf The $\chi^2$ profile for the LHC jet (left) or dijet (right) data only as $\alpha_S(M_Z^2)$ is scanned from 0.112 to 0.122, comparing the PDF fits at NNLO, aN$^{3}$LO and aN$^{3}$LO with the K-factors fixed at the values corresponding to the global best fit for $\alpha_S(M_Z^2)$.}
\label{fig:chi2_profiles_total_NNLOaN3LO_jetsvsdijetsonly}
\end{center}
\end{figure}

\begin{figure}
\begin{center}
\includegraphics[scale=0.21]{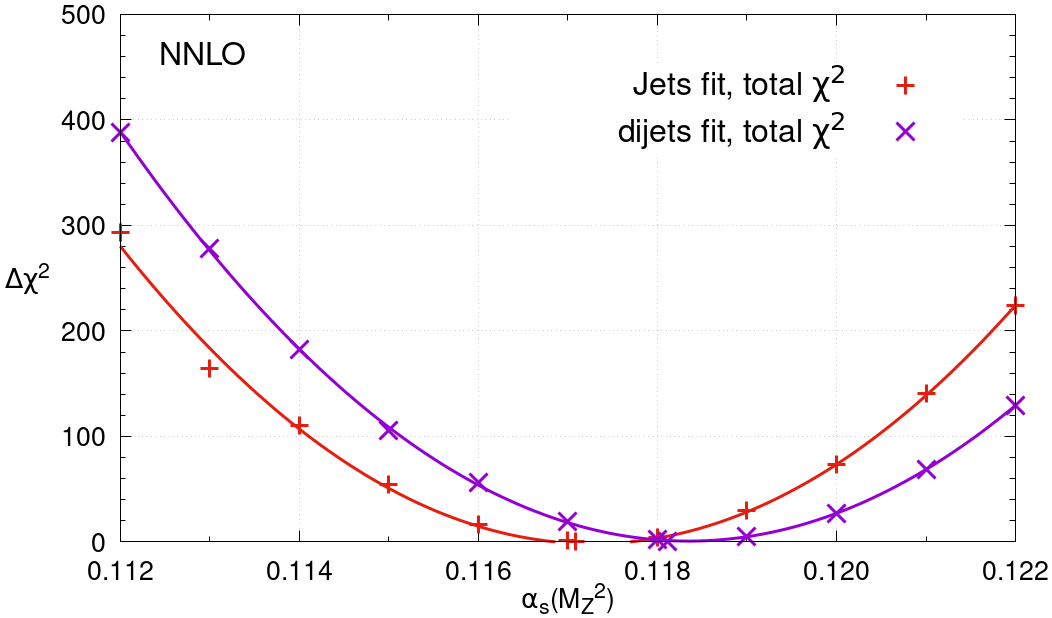}
\includegraphics[scale=0.21]{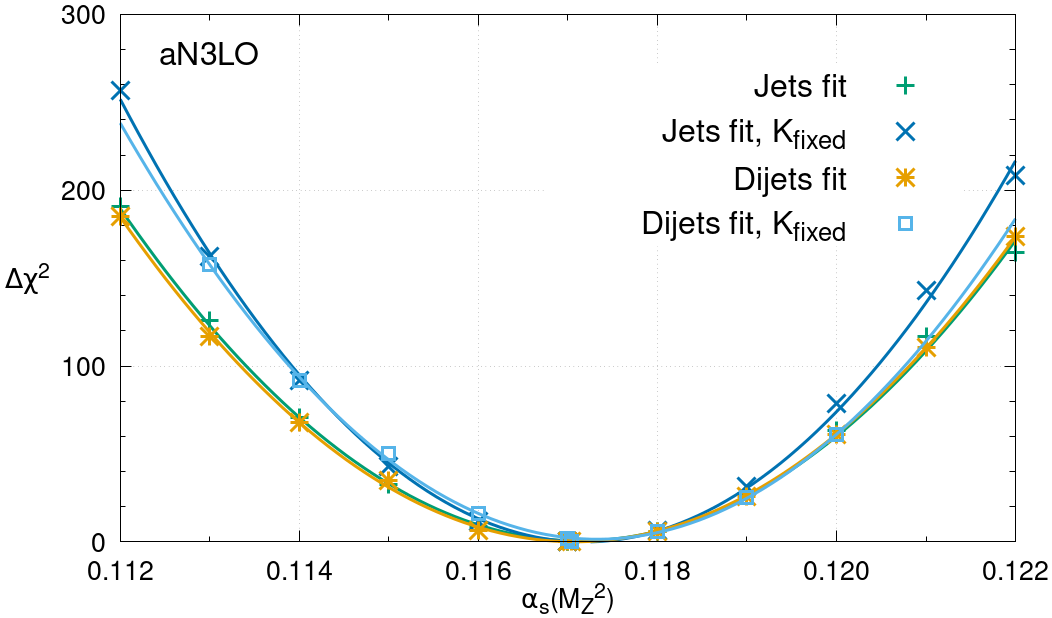}
\caption{\sf The $\chi^2$ profile for the NNLO (left) and aN$^{3}$LO (right) PDF fits as $\alpha_S(M_Z^2)$ is scanned from 0.112 to 0.122, comparing the global total $\chi^2$ profiles for the LHC inclusive jets fit with that including instead the dijets data. The aN$^{3}$LO plot also shows the effect of fixing the fitted aN$^{3}$LO k-factors to the values corresponding to the global best fit for $\alpha_S(M_Z^2)$.}
\label{fig:chi2_profiles_total_NNLOaN3LO_jetsvsdijets}
\end{center}
\end{figure}

In Fig.~\ref{fig:chi2_profiles_total_NNLOaN3LO_jetsvsdijetsonly} we show the local $\Delta \chi^2$ for the total LHC jet and dijet dataset in their respective fits, while in Fig.~\ref{fig:chi2_profiles_total_NNLOaN3LO_jetsvsdijets} we show the corresponding global $\Delta \chi^2$ profiles. At NNLO, we can see that jet data show a distinct preference for rather lower value of $\alpha_S(M_Z^2)$, with a minimum at $\sim 0.113-0.114$.  In the dijet case, on the other hand, there is a preference for a higher value,  with a minimum around $\sim 0.120$.

At aN$^{3}$LO, the situation is rather similar in the jet case; this picture (and that at NNLO) is broadly consistent with the individual breakdown shown in Fig.~\ref{fig:individual_jet_chisqs} in Section~\ref{subsec:individualdatasets}, where all datasets other than the CMS 8 TeV jets favour such a low value.
However, it is distinctly different for the dijet fit, for which the minimum now lies around $\sim 0.113$, i.e. significantly lower than at NNLO. One potential cause for this difference in behaviour is that at aN$^{3}$LO the hadronic K--factors at this order are allowed to vary in the fit, guided by predetermined priors centred at zero, see~\cite{McGowan:2022nag} for further details. As the $\alpha_S$ dependence in the local fit qualities of Fig.~\ref{fig:chi2_profiles_total_NNLOaN3LO_jetsvsdijetsonly} is to some extent induced by the explicit $\alpha_S$ dependence of the corresponding hadronic K--factors, there will be some correlation with the variation in the aN$^{3}$LO K--factors. This effect has been noted in previous sections for different datasets, in the context of the comparison between NNLO and the \anlo fits, and here we present a somewhat more detailed comparison.

Upon inspection we find that the $a_{\rm NLO}$ K--factor parameter, which contains the dominant $\alpha_S$ dependence, is directly anti--correlated with the value of the strong coupling for both the jet and dijet fits. This freedom in the aN$^{3}$LO  K--factors may therefore lead to a modification of the preferred value of $\alpha_S$. To investigate this, we also show the aN$^{3}$LO profiles, but now with the aN$^{3}$LO K--factors fixed, for concreteness at the global best fit values of $\alpha_S(M_Z^2)$. We can see that indeed the preferred value of $\alpha_S(M_Z^2)$ is now higher than with the K--factors free, but lower than at NNLO, with a minimum at around $\sim 0.117$. In the jet case, on the other hand, the result is roughly unchanged.

Therefore, the freedom in the aN$^{3}$LO K--factors does indeed induce some change in the preferred value of the strong coupling in the dijet fit, but there remains a further change due to the overall effect of working at this order. We note that while the values of the minima for the three cases in Fig.~\ref{fig:chi2_profiles_total_NNLOaN3LO_jetsvsdijetsonly} (right) are significantly  different, the corresponding $\chi^2$ profiles are rather shallow. Indeed, evaluating the corresponding confidence limits according to the hypothesis testing criteria applied in the MSHT dynamic tolerance procedure, for the aN$^{3}$LO dijet fit, the local $\chi^2$ minimum is at $\sim 0.113$, but with the 68\% C.L. region covering $\sim 0.108 - 0.118$. For the NNLO dijet fit, the local $\chi^2$ minimum of the dijet data is at $\sim 0.120$, but with the 68\% C.L. region for this data covering $\sim 0.116-0.124$. For the aN$^{3}$LO dijet fit, with the dijet aN$^{3}$LO K--factors fit to the $\alpha_S(M_Z^2)=0.118$ best fit values, the local $\chi^2$ minimum for the dijet data is at $\sim 0.117$, but with the 68\% C.L. region for this data covering $\sim 0.114 - 0.120$. Therefore, the preferred values of the strong coupling are broadly consistent within their uncertainties. We can see in particular that the freedom in the aN$^{3}$LO  K--factors, and their correlation with the value of $\alpha_S(M_Z^2)$, leads to a shallower $\chi^2$ profile as $\as$ is changed and therefore to an increase in the size of the uncertainty, with respect to the  aN$^{3}$LO  case with the dijet K--factor fixed, but also compared to the NNLO case.

In Fig.~\ref{fig:chi2_profiles_total_NNLOaN3LO_jetsvsdijets} we  show the corresponding global $\chi^2$ profiles. We note that the corresponding inclusive jet profiles are identical, by construction, to those shown in Fig.~\ref{fig:chi2_profiles_total_NNLOaN3LO}. We can see that at NNLO, the preferred value of the strong coupling is rather lower in the jet case in comparison to the dijet. This is consistent with the local fit qualities discussed above, as well as qualitatively  with the CMS analyses of jet~\cite{CMS:2014qtp,CMS:2016lna} and dijet~\cite{CMS:2017jfq} data, although as these are performed at NLO it is difficult to draw firm comparisons.
At aN$^{3}$LO, on the other hand, we can see that the preferred value of the strong coupling is now remarkably similar between the jet and dijet fits.

To be precise, the dijet best fit and uncertainties are given, after suitable symmetrising, by:
\begin{equation*}
    \as({\rm Dijet,\, NNLO}) = 0.1181 \pm 0.0012
\end{equation*}
\begin{equation*}
        \as({\rm Dijet,\, aN^{3}LO}) = 0.1170 \pm 0.0013
\end{equation*}
where the uncertainty is calculated using the usual dynamic tolerance procedure described in Section~\ref{subsec:bounds}. At NNLO, the lower bound is again set by the ATLAS 8~TeV $Z$ data\cite{ATLAS8Z3D} and the upper bound by the BCDMS $F_2$ proton data~\cite{Benvenuti:1989rh}. At \anlo, the lowest bound is set by the SLAC $F_2$ deuteron data\cite{Whitlow:1990gk,Whitlow:1991uw} and the upper by the BCDMS $F_2$ proton data~\cite{Benvenuti:1989rh}. Therefore, at both orders exactly the same datasets end up placing the most limiting bound as in the jet fit. Indeed, the only dijet dataset to place any  relevant constraint is as expected the CMS 8 TeV dijets~\cite{CMS:2017jfq}, which places a lower bound of $-0.0023$ at NNLO and an upper bound of $+0.0017$ at \anlo. The fact that a lower bound is placed at NNLO and an upper bound at \anlo is consistent with the difference in trends in the local $\chi^2$ profiles shown in Fig.~\ref{fig:chi2_profiles_total_NNLOaN3LO_jetsvsdijets}.

At NNLO, the extracted value of $\as$ is therefore $\sim 0.001$ lower in the jet fit, but these are fully consistent with each other within their quoted uncertainties, after applying the appropriate dynamic tolerance procedure. Nonetheless, it is clear that the choice of jet dataset does have a non--negligible impact on the strong coupling extraction at this order. In both cases though it remains consistent with the world average value $\alpha_S(M_Z^2)=0.1180\pm 0.0009$~\cite{Huston:2023ofk}. In fact, this again emphasises the importance of applying a tolerance in such cases, as with a $\Delta\chi^2=1$ criterion the uncertainties of the $\as$ extractions  reduce by a factor of $\sim 3$, i.e. in such a way that the NNLO inclusive jet and dijet $\as$ determinations would no longer overlap within their uncertainties.

We also note that the uncertainty bands on the strong coupling are moderately smaller in the dijet fits, at both orders. As discussed in~\cite{Cridge:2023ozx}, there are various indications that the fit to the dijet data is more stable than and hence may be preferable to the inclusive jet fit. These reduced uncertainties can be taken as further evidence of this.


At aN$^{3}$LO, on the other hand, we can see that the preferred value of the strong coupling is now remarkably similar between the jet and dijet fits, and consistent with the value of $\sim 0.1170$ found in the MSHTaN$^{3}$LO fit~\cite{McGowan:2022nag}, as we would expect in the jet case, given the similarity in the underlying datasets. Therefore, by going to this order the consistency between the two fits in terms of the preferred value of the strong coupling is improved. This is provides further evidence in support of the aN$^{3}$LO fit, and its superiority with respect to the NNLO case. 

We finally remark that the global $\chi^2$ profiles are observed to be somewhat shallower in the aN$^{3}$LO case in comparison to NNLO. This again indicates, as discussed above, that at this order the final uncertainty on the strong coupling derived from the aN$^{3}$LO may increase mildly in comparison to NNLO; while the precision may be less, the accuracy is on the other hand  improved, due to the more accurate theory in the aN$^{3}$LO fit. Moreover an additional uncertainty is now included due to missing higher order theory information. In order to assess the impact of this, one could fix the K-factors at their NNLO values and repeat the analysis. However, given it is somewhat artificial to fix the K-factors at their NNLO values in this way, we choose not to quote corresponding uncertainties calculated due to the dynamic tolerance criteria here, but we have confirmed that the corresponding uncertainty is moderately smaller, and more in line with the NNLO case, for the baseline jet fit.

\section{PDF and Cross Section Results}

\begin{figure}
\begin{center}
\includegraphics[scale=0.205]{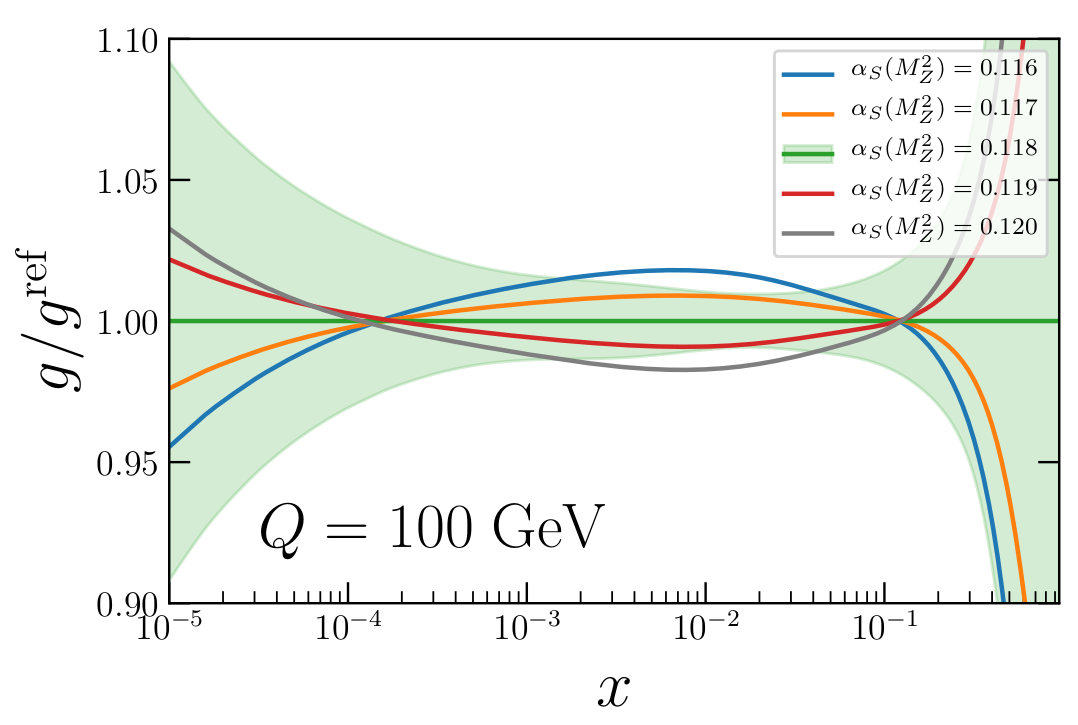}
\includegraphics[scale=0.205]{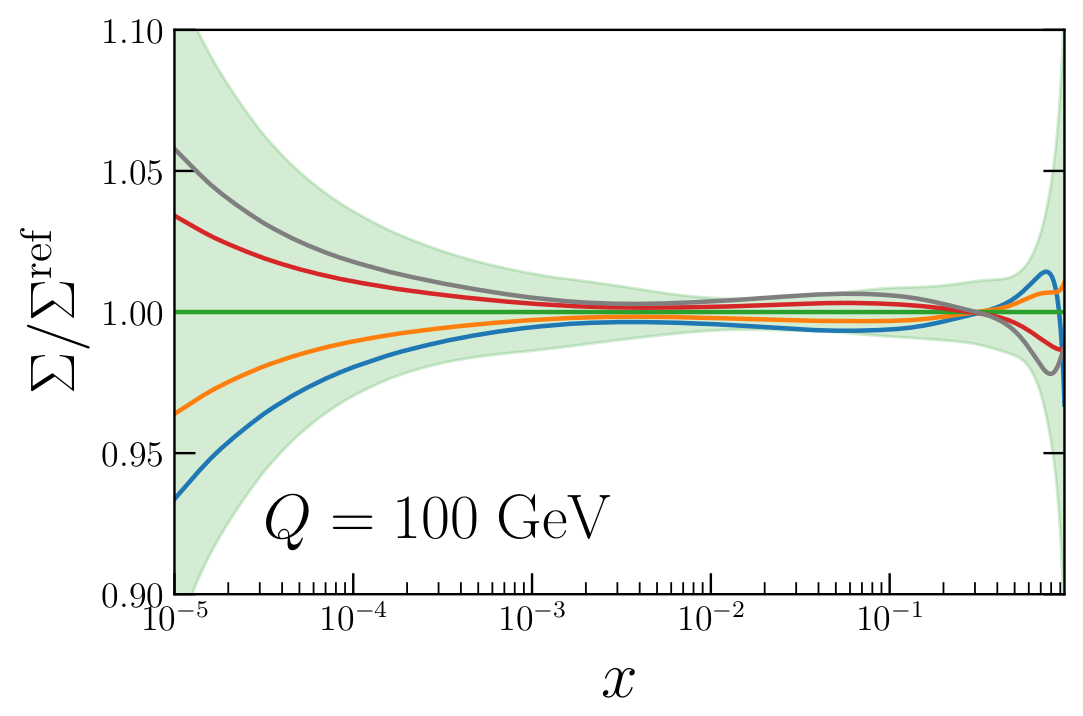}
\caption{\sf The impact of varying $\as$ in the aN$^3$LO PDF fit on (left) the extracted gluon and (right) total singlet PDFs. This demonstrates the correlations between the PDFs and $\as$.}
\label{fig:aN3LO_PDFalphas_corrs}    
\end{center} 
\end{figure}

In this section we first present the impacts of varying $\as$ on the PDFs themselves in Fig.~\ref{fig:aN3LO_PDFalphas_corrs}. In turn we utilise the PDF eigenvectors at our default fixed $\as=0.118$ in the usual Hessian manner to determine a PDF uncertainty on various inclusive LHC cross-sections which may then be combined in quadrature with the $\as$ uncertainty, determined as described below.

Beginning first with the impact of varying $\as$ on the PDFs, Fig.~\ref{fig:aN3LO_PDFalphas_corrs} shows the change of the gluon (left) and total singlet (right) aN$^3$LO PDFs as $\as$ is altered in steps of 0.001. At each new fixed value of $\as$ the PDFs are refit, as required\cite{Forte:2020pyp}. As expected, for the gluon PDF we observe a significant correlation with $\as$. Structure function data largely constrain the gluon in the intermediate to low $x$ region, as a result the fit maintains $dF_2/dQ^2 \sim \alpha_S g$, where $g$ is the gluon PDF. This therefore anti-correlates the gluon and $\as$ for $x \lesssim 0.1$. The momentum sum rule then indirectly results in a correlation between the gluon and $\as$ at high $x \gtrsim 0.1$. This is as observed at NNLO in \cite{Cridge:2021qfd}. The behaviour of the quarks is somewhat different, as illustrated by the total singlet $\Sigma(x,Q^2) = \sum_{i=1}^{N_f}(q_i(x,Q^2)+\bar{q}_i(x,Q^2))$. At large $x \gtrsim 0.3$ the singlet reduces with $\as$ due to the increased QCD splitting which depletes the quarks at large $x$. As a result however the quarks are enlarged at lower $x$, such that $\as$ and $\Sigma$ are correlated below $x \sim 0.2$, the impact though is smaller than observed for the gluon. The impacts of these changes of $\Delta \as = \pm 0.001$ are within the uncertainty bands for both the singlet (and quarks more generally - not shown) and the gluon. At lower scales, the PDF changes with $\as$ are found to be larger~\cite{Cridge:2021qfd}.

\begin{figure}
\begin{center}
\includegraphics[scale=0.66]{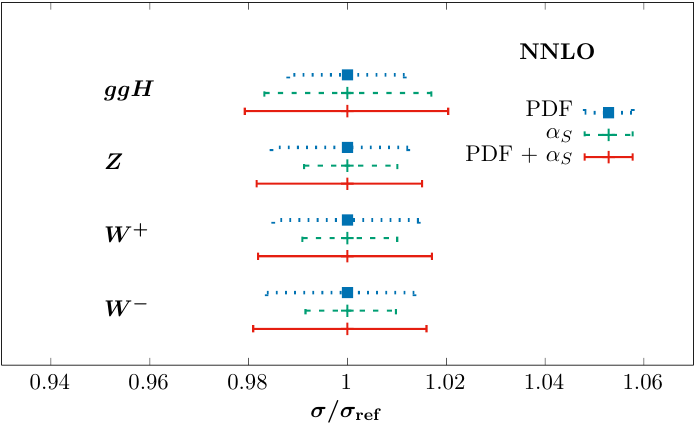}
\includegraphics[scale=0.66]{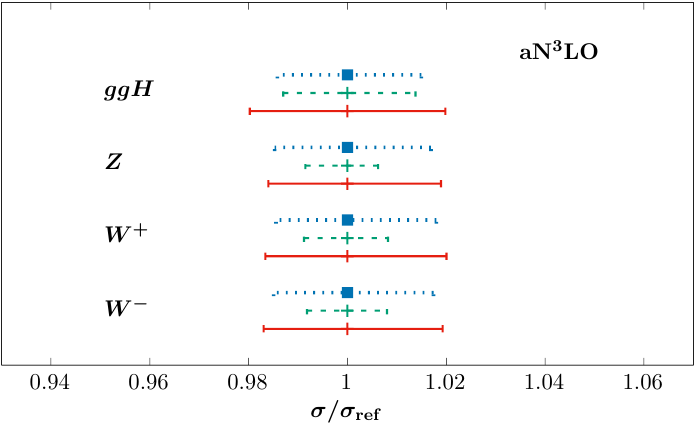}
\caption{\sf Cross section uncertainties for gluon fusion Higgs, $Z$, and $W^{\pm}$ production at $\sqrt{s}=14$~TeV at (left) NNLO with the MSHT20nnlo PDFs and (right) N$^{3}$LO with the MSHT20aN3LO PDFs. The blue dotted bars are the PDF uncertainties, the green dashed represent the $\alpha_S$ uncertainty, and the red solid bars are the combined PDF+$\alpha_S$ uncertainty, added in quadrature.}
\label{fig:aN3LO_crosssections_withalphasuncs}
\end{center}
\end{figure}

The correlations between $\as$ and the PDF central values mean that $\as$ uncertainties on cross-sections may be altered relative to the expected direct impact of $\as$ on the cross-sections due to the indirect impact on the PDFs. In Fig.~\ref{fig:aN3LO_crosssections_withalphasuncs} we show results for PDF and $\alpha_S$ uncertainties for a selection of LHC ($\sqrt{s}=14$ TeV) cross sections, namely Higgs boson production via gluon fusion, and weak boson ($W^\pm$, $Z$) production in the Drell Yan process. These are calculated using the 
\texttt{n3loxs} code~\cite{Baglio:2022wzu}, and with  the same settings as are used to calculate the cross section results shown in~\cite{Cridge:2023ryv}. For the $\alpha_S$ uncertainty we take a range of $\pm 0.001$ around the best fit value. For other variations close to this a linear scaling of the change in the prediction with $\alpha_S$ may be taken to good approximation. 

The overall trend at NNLO is very similar to the results shown in~\cite{Cridge:2021qfd} for $\sqrt{s}=13$ TeV. For the ggH cross section, the direct sensitivity to the value of $\alpha_S$ is somewhat compensated for by the anti--correlation between this and the gluon PDF,
while for $W,Z$ production the direct sensitivity to the value of $\alpha_S$ is small, and the majority of the corresponding uncertainty comes from the PDF change in the fit. The overall trends are observed to be rather similar between the NNLO and \anlo cases, with some small differences observed. For example, in the ggH case the PDF uncertainty is somewhat larger at \anlo (as observed in~\cite{McGowan:2022nag}), and thus the relative breakdown between the PDF and $\alpha_S$ uncertainty is slightly different.

\section{Conclusions}\label{sec:summary}

In this article we have studied the optimal value and uncertainty of the strong coupling resulting from the first extraction of approximate N$^3$LO PDFs made by us in \cite{McGowan:2022nag}, as well as investigating the sensitivity to using dijet rather than inclusive jet data at both NNLO and at aN$^3$LO. Our main result is that at aN$^3$LO we find that (for the default global fit including the inclusive jets data):
\begin{equation*}
        \as({\rm aN^{3}LO}) = 0.1170 \pm 0.0016.
\end{equation*}
This is in excellent agreement with the value obtained at NNLO, as well as the world average~\cite{Huston:2023ofk}, but with a slightly larger uncertainty. This might seem surprising, given that usually the 
uncertainty on $\alpha_S(M_Z^2)$ decreases with increasing perturbative order. However, the aN$^3$LO extraction is the first which correctly incorporates a theoretical uncertainty - our NNLO and NLO extractions have implicitly only included the uncertainty directly resulting from the uncertainty on the data in the PDF fit. Hence, the aN$^3$LO uncertainty is more realistic. 

We have already made the PDFs at aN$^3$LO available for a range 
of $\alpha_S(M_Z^2)$ in \cite{McGowan:2022nag}. The PDFs, can be 
obtained in \texttt{LHAPDF} format~\cite{Buckley:2014ana} at:
\\
\\
\href{http://lhapdf.hepforge.org/}{\texttt{http://lhapdf.hepforge.org/}}
\\
\\
\noindent as well as on the repository:
\\
\\ 
\href{http://www.hep.ucl.ac.uk/msht/}{\texttt{http://www.hep.ucl.ac.uk/msht/}}.
\\

The PDFs are available from $\alpha_S(M_Z^2)=0.114-0.120$ in steps of $0.001$. 
We note that these PDFs are not absolutely identical to those in this article due to a few minor corrections in the analysis and the inclusion of the ATLAS 8~TeV inclusive jet data in this article, but any differences at each value 
of $\alpha_S(M_Z^2)$ are  minor. 

The results of using the dijet rather than inclusive jet data in the analysis lead to a very good level of consistency. At NNLO the dijet analysis gives 
$\alpha_S(M_Z^2)=0.1181\pm 0.0012$, which differs from the result using inclusive jets by less than a standard deviation. At aN$^3$LO we obtain 
$\alpha_S(M_Z^2)=0.1170\pm 0.0013$, which is in almost perfect agreement with the value obtained using our default choice of inclusive jets. Hence, at NNLO and particularly at aN$^3$LO we can be confident that our extraction of the best fit value of $\alpha_S(M_Z^2)$ is reliable, and not significantly affected by our choice of input data set. There is some potential sensitivity to the unknown 
information currently missing within the approximate N$^3$LO approach. However, this is accounted for within our theoretical uncertainty. Moreover, since 
our original study took place and our framework was established, which we essentially use in this article, more information about N$^3$LO splitting functions and transition matrix elements has become available, as discussed in Section 2. We have not yet made a full study including this new information, but have performed some preliminary studies, and see no very significant effects on the fit quality and PDFs extracted. In addition we have confirmed that using the additional splitting function information now available results in an $\as$ value very close to our quoted  best fit $\as$.

Finally we note that previously there has been a tendency for the value 
of $\alpha_S(M_Z^2)$ determined from PDF fits to fall with increasing perturbative order. We do indeed still see such a trend in going from NNLO to aN$^3$LO. However, even in the case where we use dijets, where the NNLO value 
of $\alpha_S(M_Z^2)$ is higher, the change in going to aN$^3$LO is within uncertainties (unlike the situation in going from NLO to NNLO 
\cite{Cridge:2021qfd}), and when using inclusive jets there is almost no change at all. Hence, it appears as though at aN$^3$LO 
we are reaching the order at which convergence in the determined $\as$ value has essentially been achieved.

\section*{Acknowledgements}

We thank Jamie McGowan, whose invaluable work on the original a${\rm N}^3$LO fit provided the groundwork for this study. We thank the NNPDF collaboration and Emmanuele Nocera in particular for providing NLO theory grids for certain jet and dijet data sets. We thank Alex Huss for providing NNLO K-factors for the jet and dijet data. We thank Engin Eren and Katerina Lipka for help with CMS data, in particular the statistical correlations for the 8 TeV inclusive jet data. We thank Klaus Rabbertz for helpful correspondance about the experimental error definitions, and their signs, in the CMS 8 TeV dijet data.
TC acknowledges that this project has received funding from the European Research Council (ERC) under the European Union’s Horizon 2020 research and innovation programme (Grant agreement No. 101002090 COLORFREE).  L. H.-L. and R.S.T. thank STFC for support via grant awards ST/T000856/1 and ST/X000516/1.

\newpage

\bibliography{references}{}

\begin{thebibliography}{100}

\bibitem{Bailey:2020ooq}
S.~Bailey, T.~Cridge, L.~A. Harland-Lang, A.~D. Martin, and R.~S. Thorne,
\newblock Eur. Phys. J. C {\bf 81}, 341 (2021), 2012.04684.

\bibitem{Hou_2021}
T.-J. Hou {\em et~al.},
\newblock Physical Review D {\bf 103} (2021).

\bibitem{NNPDF:2021njg}
NNPDF, R.~D. Ball {\em et~al.},
\newblock Eur. Phys. J. C {\bf 82}, 428 (2022), 2109.02653.

\bibitem{ATLAS:2021vod}
ATLAS, G.~Aad {\em et~al.},
\newblock Eur. Phys. J. C {\bf 82}, 438 (2022), 2112.11266.

\bibitem{ABMP16}
S.~Alekhin, J.~Blümlein, S.~Moch, and R.~Placakyte,
\newblock Phys. Rev. D {\bf 96}, 014011 (2017), 1701.05838.

\bibitem{Vermaseren:2005qc}
J.~A.~M. Vermaseren, A.~Vogt, and S.~Moch,
\newblock Nucl. Phys. B {\bf 724}, 3 (2005), hep-ph/0504242.

\bibitem{Moch:2017uml}
S.~Moch, B.~Ruijl, T.~Ueda, J.~A.~M. Vermaseren, and A.~Vogt,
\newblock JHEP {\bf 10}, 041 (2017), 1707.08315.

\bibitem{Moch:2021qrk}
S.~Moch, B.~Ruijl, T.~Ueda, J.~A.~M. Vermaseren, and A.~Vogt,
\newblock Phys. Lett. B {\bf 825}, 136853 (2022), 2111.15561.

\bibitem{Falcioni:2023luc}
G.~Falcioni, F.~Herzog, S.~Moch, and A.~Vogt,
\newblock Phys. Lett. B {\bf 842}, 137944 (2023), 2302.07593.

\bibitem{Falcioni:2023vqq}
G.~Falcioni, F.~Herzog, S.~Moch, and A.~Vogt,
\newblock Phys. Lett. B {\bf 846}, 138215 (2023), 2307.04158.

\bibitem{Moch:2023tdj}
S.~Moch, B.~Ruijl, T.~Ueda, J.~Vermaseren, and A.~Vogt,
\newblock Phys. Lett. B {\bf 849}, 138468 (2024), 2310.05744.

\bibitem{Falcioni:2023tzp}
G.~Falcioni, F.~Herzog, S.~Moch, J.~Vermaseren, and A.~Vogt,
\newblock Phys. Lett. B {\bf 848}, 138351 (2024), 2310.01245.

\bibitem{Gehrmann:2023cqm}
T.~Gehrmann, A.~von Manteuffel, V.~Sotnikov, and T.-Z. Yang,
\newblock JHEP {\bf 01}, 029 (2024), 2308.07958.

\bibitem{Fadin:1975cb}
V.~S. Fadin, E.~A. Kuraev, and L.~N. Lipatov,
\newblock Phys. Lett. B {\bf 60}, 50 (1975).

\bibitem{Kuraev:1976ge}
E.~A. Kuraev, L.~N. Lipatov, and V.~S. Fadin,
\newblock Sov. Phys. JETP {\bf 44}, 443 (1976).

\bibitem{Lipatov:1976zz}
L.~N. Lipatov,
\newblock Sov. J. Nucl. Phys. {\bf 23}, 338 (1976).

\bibitem{Kuraev:1977fs}
E.~A. Kuraev, L.~N. Lipatov, and V.~S. Fadin,
\newblock Sov. Phys. JETP {\bf 45}, 199 (1977).

\bibitem{Fadin:1998py}
V.~S. Fadin and L.~N. Lipatov,
\newblock Phys. Lett. B {\bf 429}, 127 (1998), hep-ph/9802290.

\bibitem{Jaroszewicz:1982gr}
T.~Jaroszewicz,
\newblock Phys. Lett. B {\bf 116}, 291 (1982).

\bibitem{Ciafaloni:1998gs}
M.~Ciafaloni and G.~Camici,
\newblock Phys. Lett. B {\bf 430}, 349 (1998), hep-ph/9803389.

\bibitem{Catani:1994sq}
S.~Catani and F.~Hautmann,
\newblock Nucl. Phys. B {\bf 427}, 475 (1994), hep-ph/9405388.

\bibitem{Davies:2022ofz}
J.~Davies, C.~H. Kom, S.~Moch, and A.~Vogt,
\newblock JHEP {\bf 08}, 135 (2022), 2202.10362.

\bibitem{Kawamura:2012cr}
H.~Kawamura, N.~A. Lo~Presti, S.~Moch, and A.~Vogt,
\newblock Nucl. Phys. B {\bf 864}, 399 (2012), 1205.5727.

\bibitem{Bierenbaum:2009mv}
I.~Bierenbaum, J.~Blumlein, and S.~Klein,
\newblock Nucl. Phys. B {\bf 820}, 417 (2009), 0904.3563.

\bibitem{Ablinger:2014vwa}
J.~Ablinger {\em et~al.},
\newblock Nucl. Phys. B {\bf 886}, 733 (2014), 1406.4654.

\bibitem{Ablinger:2014nga}
J.~Ablinger {\em et~al.},
\newblock Nucl. Phys. B {\bf 890}, 48 (2014), 1409.1135.

\bibitem{Blumlein:2021enk}
J.~Bl\"umlein, P.~Marquard, C.~Schneider, and K.~Sch\"onwald,
\newblock Nucl. Phys. B {\bf 971}, 115542 (2021), 2107.06267.

\bibitem{Ablinger:2014uka}
J.~Ablinger {\em et~al.},
\newblock Nucl. Phys. B {\bf 885}, 280 (2014), 1405.4259.

\bibitem{Ablinger:2014tla}
J.~Ablinger {\em et~al.},
\newblock Nucl. Part. Phys. Proc. {\bf 258-259}, 37 (2015), 1409.1435.

\bibitem{ablinger:agq}
J.~{Ablinger} {\em et~al.},
\newblock Nuclear Physics B {\bf 882}, 263 (2014), 1402.0359.

\bibitem{Ablinger:2022wbb}
J.~Ablinger {\em et~al.},
\newblock JHEP {\bf 12}, 134 (2022), 2211.05462.

\bibitem{Ablinger:2023ahe}
J.~Ablinger {\em et~al.},
\newblock Nucl. Phys. B {\bf 999}, 116427 (2024), 2311.00644.

\bibitem{Ablinger:2024xtt}
J.~Ablinger {\em et~al.},
\newblock (2024), 2403.00513.

\bibitem{McGowan:2022nag}
J.~McGowan, T.~Cridge, L.~A. Harland-Lang, and R.~S. Thorne,
\newblock Eur. Phys. J. C {\bf 83}, 185 (2023), 2207.04739.

\bibitem{NNPDF:2024nan}
NNPDF, R.~D. Ball {\em et~al.},
\newblock (2024), 2402.18635.

\bibitem{vanRitbergen:1997va}
T.~van Ritbergen, J.~A.~M. Vermaseren, and S.~A. Larin,
\newblock Phys. Lett. B {\bf 400}, 379 (1997), hep-ph/9701390.

\bibitem{Cridge:2023ozx}
T.~Cridge, L.~A. Harland-Lang, and R.~S. Thorne,
\newblock (2023), 2312.12505.

\bibitem{Cridge:2023ryv}
T.~Cridge, L.~A. Harland-Lang, and R.~S. Thorne,
\newblock (2023), 2312.07665.

\bibitem{Cridge:2021qfd}
T.~Cridge, L.~A. Harland-Lang, A.~D. Martin, and R.~S. Thorne,
\newblock Eur. Phys. J. C {\bf 81}, 744 (2021), 2106.10289.

\bibitem{ATLAS:2017kux}
ATLAS, M.~Aaboud {\em et~al.},
\newblock JHEP {\bf 09}, 020 (2017), 1706.03192.

\bibitem{Huston:2023ofk}
J.~Huston, K.~Rabbertz, and G.~Zanderighi,
\newblock (2023), 2312.14015.

\bibitem{dEnterria:2022hzv}
D.~d'Enterria {\em et~al.},
\newblock (2022), 2203.08271.

\bibitem{Forte:2020pyp}
S.~Forte and Z.~Kassabov,
\newblock Eur. Phys. J. C {\bf 80}, 182 (2020), 2001.04986.

\bibitem{Benvenuti:1989rh}
BCDMS, A.~C. Benvenuti {\em et~al.},
\newblock Phys. Lett. {\bf B223}, 485 (1989).

\bibitem{Arneodo:1996qe}
New Muon, M.~Arneodo {\em et~al.},
\newblock Nucl. Phys. {\bf B483}, 3 (1997), hep-ph/9610231.

\bibitem{Whitlow:1991uw}
L.~W. Whitlow, E.~M. Riordan, S.~Dasu, S.~Rock, and A.~Bodek,
\newblock Phys. Lett. B {\bf 282}, 475 (1992).

\bibitem{Whitlow:1990gk}
L.~W. Whitlow, S.~Rock, A.~Bodek, E.~M. Riordan, and S.~Dasu,
\newblock Phys. Lett. B {\bf 250}, 193 (1990).

\bibitem{Martin:1998sq}
A.~D. Martin, R.~G. Roberts, W.~J. Stirling, and R.~S. Thorne,
\newblock Eur. Phys. J. C {\bf 4}, 463 (1998), hep-ph/9803445.

\bibitem{ATLASWZ7f}
ATLAS, M.~Aaboud {\em et~al.},
\newblock Eur. Phys. J. C {\bf 77}, 367 (2017), 1612.03016.

\bibitem{ATLASW8}
ATLAS, G.~Aad {\em et~al.},
\newblock Eur. Phys. J. C {\bf 79}, 760 (2019), 1904.05631.

\bibitem{CMSW8}
CMS, V.~Khachatryan {\em et~al.},
\newblock Eur. Phys. J. C {\bf 76}, 469 (2016), 1603.01803.

\bibitem{LHCbZ7}
LHCb, R.~Aaij {\em et~al.},
\newblock JHEP {\bf 08}, 039 (2015), 1505.07024.

\bibitem{LHCbWZ8}
LHCb, R.~Aaij {\em et~al.},
\newblock JHEP {\bf 01}, 155 (2016), 1511.08039.

\bibitem{Caola:2022ayt}
F.~Caola {\em et~al.},
\newblock {The Path forward to N$^3$LO},
\newblock in {\em {Snowmass 2021}}, 2022, 2203.06730.

\bibitem{Duhr:2020seh}
C.~Duhr, F.~Dulat, and B.~Mistlberger,
\newblock Phys. Rev. Lett. {\bf 125}, 172001 (2020), 2001.07717.

\bibitem{Duhr:2020sdp}
C.~Duhr, F.~Dulat, and B.~Mistlberger,
\newblock JHEP {\bf 11}, 143 (2020), 2007.13313.

\bibitem{Gehrmann:DYN3LO}
X.~Chen {\em et~al.},
\newblock Phys. Rev. Lett. {\bf 128}, 052001 (2022), 2107.09085.

\bibitem{duhr:DY2021}
C.~Duhr and B.~Mistlberger,
\newblock JHEP {\bf 03}, 116 (2022), 2111.10379.

\bibitem{CMS:2014nvq}
CMS, S.~Chatrchyan {\em et~al.},
\newblock Phys. Rev. D {\bf 90}, 072006 (2014), 1406.0324.

\bibitem{ATLAS:2014riz}
ATLAS, G.~Aad {\em et~al.},
\newblock JHEP {\bf 02}, 153 (2015), 1410.8857,
\newblock [Erratum: JHEP 09, 141 (2015)].

\bibitem{CMS:2016lna}
CMS, V.~Khachatryan {\em et~al.},
\newblock JHEP {\bf 03}, 156 (2017), 1609.05331.

\bibitem{Jing:2023isu}
X.~Jing {\em et~al.},
\newblock Phys. Rev. D {\bf 108}, 034029 (2023), 2306.03918.

\bibitem{Cridge:2021qjj}
PDF4LHC21 combination group, T.~Cridge,
\newblock SciPost Phys. Proc. {\bf 8}, 101 (2022), 2108.09099.

\bibitem{PDF4LHCWorkingGroup:2022cjn}
PDF4LHC Working Group, R.~D. Ball {\em et~al.},
\newblock J. Phys. G {\bf 49}, 080501 (2022), 2203.05506.

\bibitem{Tevatron-top}
CDF, D0, T.~A. Aaltonen {\em et~al.},
\newblock Phys. Rev. {\bf D89}, 072001 (2014), 1309.7570.

\bibitem{ATLAS-top7-1}
ATLAS, G.~Aad {\em et~al.},
\newblock Eur. Phys. J. {\bf C71}, 1577 (2011), 1012.1792.

\bibitem{ATLAS-top7-2}
ATLAS, G.~Aad {\em et~al.},
\newblock Phys. Lett. {\bf B707}, 459 (2012), 1108.3699.

\bibitem{ATLAS-top7-3}
ATLAS, G.~Aad {\em et~al.},
\newblock Phys. Lett. {\bf B711}, 244 (2012), 1201.1889.

\bibitem{ATLAS-top7-4}
ATLAS, G.~Aad {\em et~al.},
\newblock JHEP {\bf 05}, 059 (2012), 1202.4892.

\bibitem{ATLAS-top7-5}
ATLAS, G.~Aad {\em et~al.},
\newblock Phys. Lett. {\bf B717}, 89 (2012), 1205.2067.

\bibitem{ATLAS-top7-6}
ATLAS, G.~Aad {\em et~al.},
\newblock Eur. Phys. J. {\bf C73}, 2328 (2013), 1211.7205.

\bibitem{CMS-top7-1}
CMS, S.~Chatrchyan {\em et~al.},
\newblock Phys. Rev. D {\bf 85}, 112007 (2012), 1203.6810.

\bibitem{CMS-top7-2}
CMS, S.~Chatrchyan {\em et~al.},
\newblock JHEP {\bf 11}, 067 (2012), 1208.2671.

\bibitem{CMS-top7-3}
CMS, S.~Chatrchyan {\em et~al.},
\newblock Phys. Lett. B {\bf 720}, 83 (2013), 1212.6682.

\bibitem{CMS-top7-4}
CMS, S.~Chatrchyan {\em et~al.},
\newblock Eur. Phys. J. C {\bf 73}, 2386 (2013), 1301.5755.

\bibitem{CMS-top7-5}
CMS, S.~Chatrchyan {\em et~al.},
\newblock JHEP {\bf 05}, 065 (2013), 1302.0508.

\bibitem{CMS-top8}
CMS, S.~Chatrchyan {\em et~al.},
\newblock JHEP {\bf 02}, 024 (2014), 1312.7582,
\newblock [Erratum: JHEP 02, 102 (2014)].

\bibitem{CMSttbar08_ytt}
CMS, V.~Khachatryan {\em et~al.},
\newblock Eur. Phys. J. C {\bf 75}, 542 (2015), 1505.04480.

\bibitem{ATLASsdtop}
ATLAS, G.~Aad {\em et~al.},
\newblock Eur. Phys. J. C {\bf 76}, 538 (2016), 1511.04716.

\bibitem{ATLASttbarDilep08_ytt}
ATLAS, M.~Aaboud {\em et~al.},
\newblock Phys. Rev. D {\bf 94}, 092003 (2016), 1607.07281.

\bibitem{Cridge:2023ztj}
T.~Cridge and M.~A. Lim,
\newblock Eur. Phys. J. C {\bf 83}, 805 (2023), 2306.14885.

\bibitem{ATLASZpT}
ATLAS, G.~Aad {\em et~al.},
\newblock Eur. Phys. J. C {\bf 76}, 291 (2016), 1512.02192.

\bibitem{atlascollaboration2023precise}
A.~Collaboration,
\newblock A precise measurement of the z-boson double-differential transverse
  momentum and rapidity distributions in the full phase space of the decay
  leptons with the atlas experiment at $\sqrt s$ = 8 tev, 2023, 2309.09318.

\bibitem{ATLAS:2023lhg}
ATLAS, G.~Aad {\em et~al.},
\newblock (2023), 2309.12986.

\bibitem{Martin:2009iq}
A.~D. Martin, W.~J. Stirling, R.~S. Thorne, and G.~Watt,
\newblock Eur. Phys. J. C {\bf 63}, 189 (2009), 0901.0002.

\bibitem{MMHTas}
L.~A. Harland-Lang, A.~D. Martin, P.~Motylinski, and R.~S. Thorne,
\newblock Eur. Phys. J. C {\bf 75}, 435 (2015), 1506.05682.

\bibitem{collins2001tests}
J.~C. Collins and J.~Pumplin,
\newblock Tests of goodness of fit to multiple data sets, 2001, hep-ph/0105207.

\bibitem{Harland_Lang_2015}
L.~A. Harland-Lang, A.~D. Martin, P.~Motylinski, and R.~S. Thorne,
\newblock The European Physical Journal C {\bf 75} (2015).

\bibitem{Schmidt_2018}
C.~Schmidt, J.~Pumplin, and C.-P. Yuan,
\newblock Physical Review D {\bf 98} (2018).

\bibitem{ATLAS8Z3D}
ATLAS, M.~Aaboud {\em et~al.},
\newblock JHEP {\bf 12}, 059 (2017), 1710.05167.

\bibitem{ATLASHMDY8}
ATLAS, G.~Aad {\em et~al.},
\newblock JHEP {\bf 08}, 009 (2016), 1606.01736.

\bibitem{Hou:2019efy}
T.-J. Hou {\em et~al.},
\newblock Phys. Rev. D {\bf 103}, 014013 (2021), 1912.10053.

\bibitem{Ball_2018}
R.~D. Ball {\em et~al.},
\newblock The European Physical Journal C {\bf 78} (2018).

\bibitem{H1+ZEUScharm}
H1, ZEUS, H.~Abramowicz {\em et~al.},
\newblock Eur. Phys. J. C {\bf 73}, 2311 (2013), 1211.1182.

\bibitem{ATLAS:2013jmu}
ATLAS, G.~Aad {\em et~al.},
\newblock JHEP {\bf 05}, 059 (2014), 1312.3524.

\bibitem{CMS:2012ftr}
CMS, S.~Chatrchyan {\em et~al.},
\newblock Phys. Rev. D {\bf 87}, 112002 (2013), 1212.6660,
\newblock [Erratum: Phys.Rev.D 87, 119902 (2013)].

\bibitem{CMS:2017jfq}
CMS, A.~M. Sirunyan {\em et~al.},
\newblock Eur. Phys. J. C {\bf 77}, 746 (2017), 1705.02628.

\bibitem{CMS:2014qtp}
CMS, V.~Khachatryan {\em et~al.},
\newblock Eur. Phys. J. C {\bf 75}, 288 (2015), 1410.6765.

\bibitem{Baglio:2022wzu}
J.~Baglio, C.~Duhr, B.~Mistlberger, and R.~Szafron,
\newblock JHEP {\bf 12}, 066 (2022), 2209.06138.

\bibitem{Buckley:2014ana}
A.~Buckley {\em et~al.},
\newblock Eur. Phys. J. C {\bf 75}, 132 (2015), 1412.7420.

\end{thebibliography}
\bibliographystyle{h-physrev}

\end{document}